\documentclass[electronics,article,accept,pdftex,moreauthors]{Definitions/mdpi} 

\firstpage{1} 
\pubvolume{1}
\issuenum{1}
\articlenumber{0}
\pubyear{2025}
\copyrightyear{2025}
\externaleditor{Firstname Lastname} 
\datereceived{8 January 2025} 
\daterevised{15 February 2025} 
\dateaccepted{18 February 2025} 
\datepublished{ } 
\hreflink{https://doi.org/} 
\usepackage[labelformat=simple]{subcaption}								
								
\DeclareCaptionLabelFormat{subcaptionlabel}{\normalfont(\textbf{#2}\normalfont)}								
\Title{Fostering Inclusion: A Virtual Reality Experience to Raise Awareness of Dyslexia-Related Barriers in University Settings}

\TitleCitation{Fostering Inclusion: A Virtual Reality Experience to Raise Awareness of Dyslexia-Related Barriers in University Settings}

\Author{José 
 Manuel Alcalde-Llergo $^{1,2,}$*\orcidA{}, Pilar Aparicio-Martínez $^{3}$\orcidC{}, Andrea Zingoni $^{2}$, Sara Pinzi $^{4}$\orcidD{} \mbox{and Enrique Yeguas-Bolívar} $^{1,}$*\orcidB{}}

\AuthorNames{José Manuel Alcalde-Llergo, Pilar Aparicio-Martínez, Andrea Zingoni, Sara Pinzi and Enrique Yeguas-Bolívar}

\AuthorCitation{Alcalde-Llergo, 
 J.M.; Aparicio-Martínez, P.; Zingoni, A.; Pinzi, S.; Yeguas-Bolívar, E.}

\address{%
$^{1}$ \quad Department of Computer Science, University of Córdoba, 14071 {Córdoba}
, Spain\\
$^{2}$ \quad Department of Economics, Engineering, Business and Society, University of Tuscia, 01100 {Viterbo}, Italy; andrea.zingoni@unitus.it 
\\
$^{3}$ \quad Department of Nursing, Physiotherapy and Pharmacology, University of Córdoba, 14071 {Córdoba}, Spain; n32apmap@uco.es\\
$^{4}$ \quad Vicerectorate for Equality and Inclusion, University of Córdoba, 14071 {Córdoba}, Spain; qf1pinps@uco.es}

\corres{Correspondence: jose.alcalde@unitus.it (J.M.A.-L.); eyeguas@uco.es (E.Y.-B.)}

\abstract{This work introduces the design, implementation, and validation of a virtual reality (VR) experience aimed at promoting the inclusion of individuals with dyslexia in university settings. Unlike traditional awareness methods, this immersive approach offers a novel way to foster empathy by allowing participants to experience firsthand the challenges faced by students with dyslexia. Specifically, the experience raises awareness by exposing non-dyslexic individuals to the difficulties commonly encountered by dyslexic students. In the virtual environment, participants explore a virtual campus with multiple buildings, navigating between them while completing tasks and simultaneously encountering barriers that simulate some of the challenges faced by individuals with dyslexia. These barriers include reading signs with shifting letters, following directional arrows that may point incorrectly, and dealing with a lack of assistance. The campus is a comprehensive model featuring both indoor and outdoor spaces and supporting various modes of locomotion. To validate the experience, more than 30 non-dyslexic participants from the university environment, mainly professors and students, evaluated it through ad hoc satisfaction surveys. The results indicated heightened awareness of the barriers encountered by students with dyslexia, with participants deeming the experience a valuable tool for increasing visibility and fostering understanding of dyslexic students.}


\begin{document}

\section{Introduction}

Digital integration into education is gradually transforming the understanding and support of learning differences, showing more and more promising results. Among~these technologies, VR and augmented reality (AR) allow the creation  of immersive environments where users can experience and engage with simulated real-world obstacles~\citep{Xie2021}. In~particular, VR technologies are increasingly included in diverse segments of society, increasing awareness of social issues. Despite its potential, research into leveraging VR for comprehension regarding learning disorders, such as dyslexia, remains in its early stages, especially in higher education settings~\citep{zingoni_investigating_2021,Benedetti2022}.

Among the diverse learning disorders, dyslexia is one of the most prevalent learning disorders, yet it often goes undiagnosed in its early stages. Dyslexia is a neurobiologically specific learning disorder (SLD) that affects cognitive and social development~\cite{Butterfuss2018,Farah2021}. Typically neurodevelopmental and often genetic, it impairs reading and language processing and is frequently accompanied by challenges in orientation, comprehension, and~time management within academic settings~\cite {miles2013dyslexia,Franceschini2022}. These barriers can profoundly impact the educational experiences of affected students, highlighting the need for innovative approaches to foster awareness and inclusion. It is estimated that dyslexia affects between 5\% and 10\% of the global population~\citep{costantini_psychosocial_2020}. However, more recent studies suggest that this figure is likely underreported, with~the actual prevalence potentially ranging between 15\% and 20\%~\citep{dyslexia_basics}. Moreover, these statistics must account for previously noted diagnostic variability, which opposes establishing a standardized and precise measurement framework~\citep{Snowling2024}.

While prior studies have explored VR~\cite{vrEducationReview} and also AR~\citep{Materazzini2024} applications in education, few have focused on its potential to portray the lived experiences of individuals with SLDs. To~effectively support students with dyslexia, it is essential to understand the hurdles in their daily lives, especially in the learning environment. This approach would allow classmates and teachers to directly understand the importance of offering support to help students overcome the barriers they encounter~\citep{Rohmer2022}. Existing research in this field underscores the importance of this understanding in creating supportive learning environments. For~instance, earlier works have demonstrated how VR can effectively simulate sensory or physical impairments, but~there is limited evidence of its application for cognitive and learning challenges~\citep{vrVisualImapirmentSimulation}. Other studies have explored the role of VR in childhood education for students with dyslexia, demonstrating its potential to enhance cognitive skills and reading abilities~\citep{Passig2011, Passig2008, Wadlington2008}. However, research on the application of immersive simulations in higher education remains scarce, particularly in addressing how such experiences can shape perceptions and foster inclusive learning environments for students with dyslexia. This gap in research provides an opportunity to explore how immersive simulations might influence perceptions of and attitudes toward students with dyslexia, increasing awareness and enabling inclusive~education. 

The purpose of this study is to design, implement, and~validate a VR experience that immerses participants in the role of a university student with dyslexia. Within~this virtual campus, participants navigate various campus buildings to complete tasks, encountering barriers that reflect the challenges associated with this SLD. Guidance is provided through a map, directional arrows, and~signs.
Through this approach, the~study aims to increase consciousness of the specific barriers students with dyslexia face and to promote awareness among non-dyslexic individuals. Key features of the experience include tasks such as interpreting shifting text on signs, navigating misleading directional arrows, and~managing high-pressure situations without help. All of these scenarios mirror real-world barriers faced by individuals with dyslexia~\citep{miles2013dyslexia}.
This work fosters the potential to enhance educational practices and increase VR's role in promoting inclusion. By~involving non-dyslexic university participants in the experience, including students and professors, the~study provides valuable insights into how simulated experiences can reshape perceptions, and it allows for a deeper understanding of dyslexia at the higher education level. Beyond~raising awareness, this initiative aims to foster the adoption of more inclusive educational strategies and improve the availability of compensatory tools for students with dyslexia. By~promoting a deeper understanding of the challenges faced by these students, the~study encourages institutions to implement supportive policies and accommodations that enhance academic engagement, attendance, and~overall attainment. Moreover, as~part of the VRAIlexia project, the~outcomes of this work contribute to a broader framework for institutional change, including the release of a Memorandum of Understanding~( 
 \url{https://vrailexia.eu/the-project/io5-outside-the-box/}, accessed on 19 February 2025), which provides concrete guidelines for fostering inclusion and improving accessibility in higher education~settings.

The paper is organized as follows: Section~\ref{sec2} reviews related works, providing a comprehensive overview of existing research on integrating VR in education and its potential to simulate the experiences of individuals with SLDs, with~a particular focus on dyslexia. Section~\ref{sec3} outlines the VR application, describing the design, development, and~implementation of the VR experience, as~well as the methodologies employed to assess its effectiveness in promoting awareness. Section~\ref{sec4} presents the results, highlighting key findings from participant interactions with the VR experience, including changes in the perception and understanding of dyslexia. Finally, Section~\ref{sec5} concludes the work by discussing the broader implications of the study, its limitations, and~opportunities for future research in both educational practices and VR-based inclusion~initiatives.

\section{Related~Works}\label{sec2}

This section reviews the literature on the use of VR for raising awareness and simulating barriers, particularly in the context of dyslexia. It examines recent studies and experiences, highlighting their results and exploring the potential of VR as a tool for fostering understanding among individuals without dyslexia in university~settings.

VR has emerged as a powerful tool in promoting inclusion by providing immersive experiences that foster empathy for and understanding of individuals with disabilities~\citep{Herrera2018}. VR allows users to experience the world from the perspective of individuals with physical or cognitive impairments, facilitating emotional and cognitive connections that can lead to greater social awareness and compassion. For~instance, VR has been employed in workshops on social inclusion, helping participants better comprehend the challenges faced by others~\citep{de_luca_virtual_2023}. A~notable example of a direct impact in this field is Meta's ``VR for Good'' program~\citep{MetaVRForGood}, which encompasses multiple virtual reality experiences designed to enhance awareness of various marginalized communities.
Moreover, VR can be used to create inclusive environments that reproduce the specific needs of individuals with disabilities. For~example, VR can simulate scenarios that replicate real-world challenges faced by people with autism, helping both individuals and society at large gain a better understanding of the condition~\citep{didehbani_virtual_2016}. Other studies also prove that VR can help people with the training and rehabilitation of people with hearing impairments and intellectual disabilities, teaching them to interact with their surroundings in new and beneficial ways, such as using visual or tactile cues in place of audio~\citep{Serafin2023}.

The application of VR to simulate the experience of physical or cognitive barriers has proven to be a valuable tool for raising awareness about people with disabilities. One of the most prominent uses of VR in this context is in the simulation of sensory impairments, such as blindness, hearing loss, or~cognitive disabilities, to~help participants understand the challenges faced by individuals with these conditions.
For example, the~VR experience “Notes on Blindness: Into Darkness”~\citep{arteNotesOnBlindness} immerses users in the experience of gradually losing their vision, using audio tracks to represent how a person with blindness perceives their environment. Similarly, studies have shown that VR simulations of the obstacles faced by wheelchair users can be effective in changing participants' attitudes toward disability and fostering a sense of empathy~\citep{hoter_effects_2022}. These VR experiences have the power to evoke a sense of perspective that might be difficult to achieve through traditional awareness-raising methods, creating lasting emotional and cognitive connections between users and those living with these challenges.
Moreover, VR has been used to simulate disorientation and cognitive overload, which can mirror the challenges faced by individuals with SLDs such as dyslexia. These simulations allow participants to experience firsthand how difficult it can be to navigate complex environments, read text, or~manage multitasking under stress, thus promoting empathy and understanding~\citep{Rohmer2022}.

Focusing on the field of dyslexia, VR has also been utilized in various studies to support dyslexic children by providing engaging, interactive environments that address their specific learning needs. For~example, the~European project FORDYSVAR has developed a VR game designed for dyslexic children aged 10--16 to help them improve their reading and language skills in a more interactive and enjoyable way~\citep{rodriguez_cano_tecnologias_2021}. Other studies have explored how VR can help children with dyslexia overcome challenges related to attention, memory, and~task-switching by offering immersive and focused learning environments that are both fun and educational~\citep{pedroli_psychometric_2017}. Similarly,~\cite{KaplanRakowski2023} highlights the increasing adoption of VR in education, reporting promising results from a large-scale study involving more than 2000~teachers.

Although these studies primarily focus on assisting individuals with dyslexia by providing adaptive learning and support mechanisms, our study is inspired by their underlying goal: improving the quality of life and educational experiences of dyslexic individuals. Rather than targeting dyslexic students themselves, our approach leverages VR to raise awareness among non-dyslexic individuals, helping them understand the cognitive barriers associated with dyslexia. This aligns with the broader objective shared by the aforementioned projects, fostering inclusivity by bridging the gap between those who experience these challenges and those who interact with them in academic and social~environments.

Finally, in~the context of higher education, VR is beginning to be applied to support adult students with dyslexia. The~VRAIlexia project is one of the first initiatives designed to assist university students with dyslexia by leveraging VR and artificial intelligence technologies~\citep{zingoni_investigating_2021}. This project aims to develop customized learning support tools using VR simulations that replicate the cognitive challenges faced by dyslexic students in a university setting, such as difficulties with reading, note-taking, and~navigating among large educational buildings.
In addition to its VR simulations, the~VRAIlexia project implements different recommendation systems to provide tailored educational interventions, ensuring that the support is personalized to the individual’s needs~\citep{zingoni_ML_2023}. This innovative approach provides valuable insights into how VR can be used as a genuine assistive technology that can help students with dyslexia succeed in higher education~\citep{AlcaldePotionsMetroxraine}. VRAIlexia similarly uses VR to administer psychometric tests without the need for expert presence, making it easier to detect conditions in individuals who might otherwise remain undiagnosed in adulthood~\citep{yeguas-bolivar_determining_2022}.

\section{VR Empathy~Experience}\label{sec3}

During this section, the~proposed experience and its implementation will be detailed. This includes an overview of the tools used for development, the~workflow followed throughout the process, the~main elements involved, the~simulation of barriers, and, finally, the~gameplay design and the available~experience. 

The experience implementation was performed using the Unity game engine~\citep{unity_2014} for deployment on Meta Quest devices~\citep{Meta}. The~software development followed the Scrum methodology and adhered to best practices, resulting in an iterative process~\citep{Schwaber2020}. This approach led to multiple versions of the virtual environment's scenarios, ensuring their quality and functionality, 
and~the effectiveness of tasks performed at various campus locations was thoroughly~validated.

\subsection{Overview}

This section provides an overview of the methodology followed in this work. 
The proposed experience in this work consists of performing a series of tasks in a virtual environment that simulates a university campus. This campus is vast and consists of numerous similar buildings, which will complicate navigation through the map. The~goal of the experience is to simulate some of the barriers faced by university students with dyslexia in their daily lives. The~design guidelines followed for the development of this environment are defined~below.

At the beginning of the experience, participants find themselves outside a university campus in front of a sign that shows them how to take their first steps within the virtual environment. Following the provided instructions, the~user will go through a brief tutorial on the various available modes of locomotion, which will be very important for navigating the entire campus. After~this, the~participant must locate a map that will give them a series of directions to ensure that they can reach the exam room on~time.

To reach their destination, they will need to complete two tasks: collecting the materials needed for the exam and finding the classroom where the exam is being held. However, during~these tasks, participants will face significant barriers similar to those that students with dyslexia encounter in their daily lives, such as difficulties with reading, challenges in orientation, and~issues with memorization. These barriers will be simulated within the experience using different strategies, such as altering the order of letters in words and providing hard-to-interpret maps, among~others, which are detailed further in Section~\ref{sec:barriers}.

If the user successfully completes the tasks on time, they will reach the exam room and fulfill their objective on the~campus.

In addition, the~objective of the virtual campus is not only to raise awareness about dyslexia in university environments but~also to serve as a link between other VR experiences that aim to address more specific aspects of dyslexia awareness and other vulnerable groups. These experiences will be incorporated into the campus as they are developed, ultimately aiming to create a comprehensive virtual environment that allows users to understand firsthand the challenges these groups~face.

Once the experience is completed, its effectiveness will be evaluated through a survey. The~questions were designed based on recommendation models tailored to support dyslexic students~\citep{Morciano2024}, integrating elements from the VR aspect of the experience. This evaluation will serve as an initial analysis of the experience's effectiveness, preparing the way for future implementations and the integration of empathy tests into the campus environment. A~more detailed analysis will then assess their impact on raising awareness about dyslexia and fostering~empathy.

\subsection{Design}

The virtual environment simulates a large-scale university campus composed of multiple interconnected buildings and pathways. This environment includes seven distinct buildings that participants must navigate as they complete the tasks outlined in the experience. The~design prioritizes inclusivity and immersion, leveraging VR to simulate real-world challenges faced by individuals with~dyslexia.

To raise awareness of dyslexia and highlight the contributions of individuals with this SLD, each building within the campus is named after renowned figures with dyslexia. This approach aims to help non-dyslexic users recognize the achievements of individuals who have successfully overcome similar challenges, fostering a deeper connection between the experience and real-world examples of resilience and success. In~addition, to~enhance the sense of presence and realism, the~environment incorporates ambient sounds and common visual elements found in real university settings, such as trees, benches, bike parkings, and~avatars representing students. These elements collectively enrich the immersive experience while emphasizing empathy for the barriers encountered by individuals with~dyslexia.

Among the virtual campus's core components, the~following elements stand out due to their significance in promoting navigation, task progression, and~overall engagement:

\begin{itemize}
    \item \textbf{{Maps} 
}: To facilitate navigation within the expansive campus, two types of maps are provided to~participants: 
    \begin{itemize}
    \item \textbf{{Vertical Map}:} this simplified map highlights only the user's current location and the target destination, enabling a focused and streamlined experience (see Figure~\ref{fig:maps}a). 
    \item \textbf{{Interactive 3D Map}:} Offering a comprehensive, overhead 3D view of the campus, this map displays the names and positions of all buildings along with the user's location. Users can interact with this map to receive real-time navigation assistance toward specific destinations by pressing a designated button (see Figure~\ref{fig:maps}b).
    \end{itemize}
        \begin{figure}[H]
            
            \begin{subfigure}[b]{0.45\textwidth}
                 \centering
                 \includegraphics[width=\textwidth]{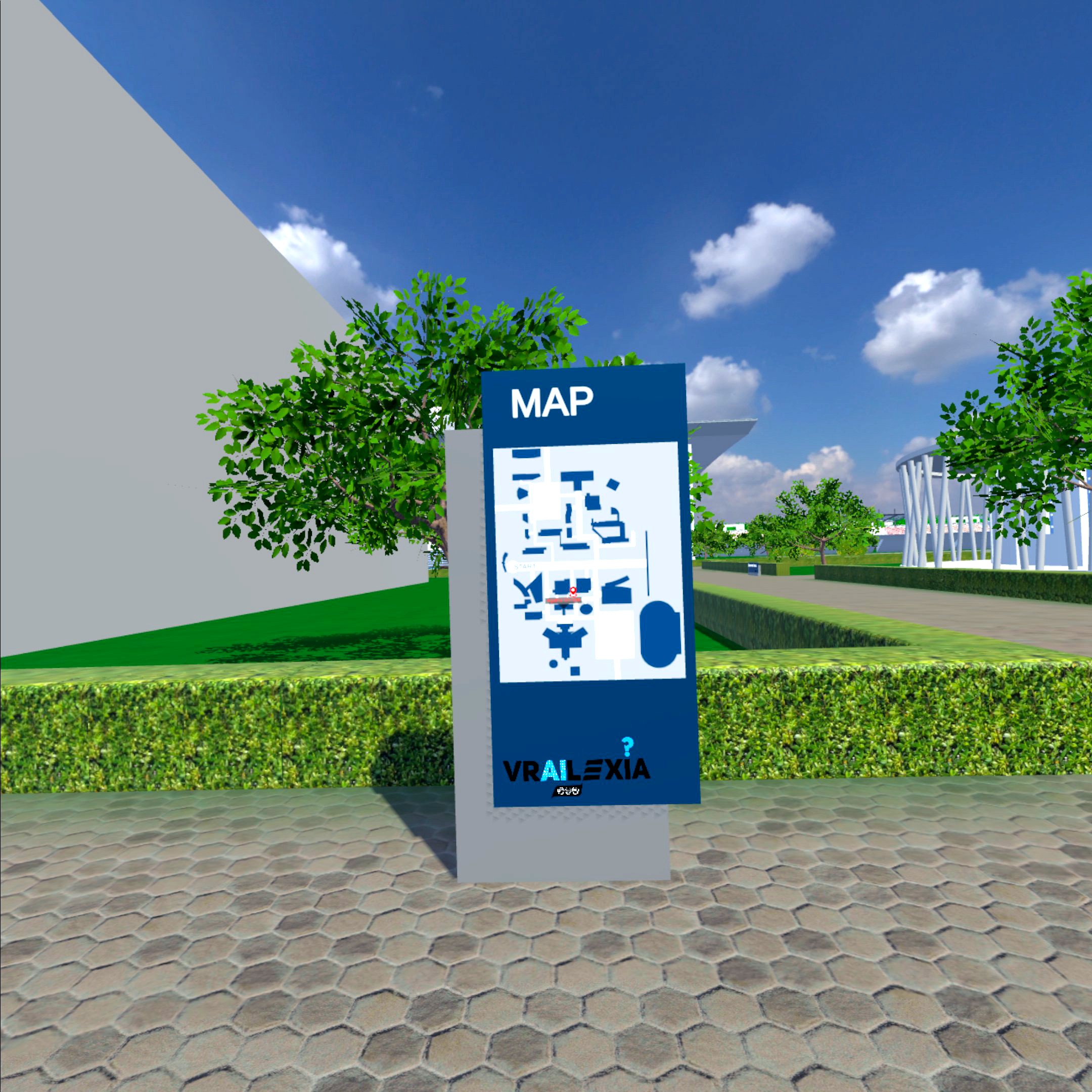}
                 \caption{\centering Vertical~Map.}
                 \label{fig:v_map}
            \end{subfigure}
            \begin{subfigure}[b]{0.45\textwidth}
                 \centering
                 \includegraphics[width=\textwidth]{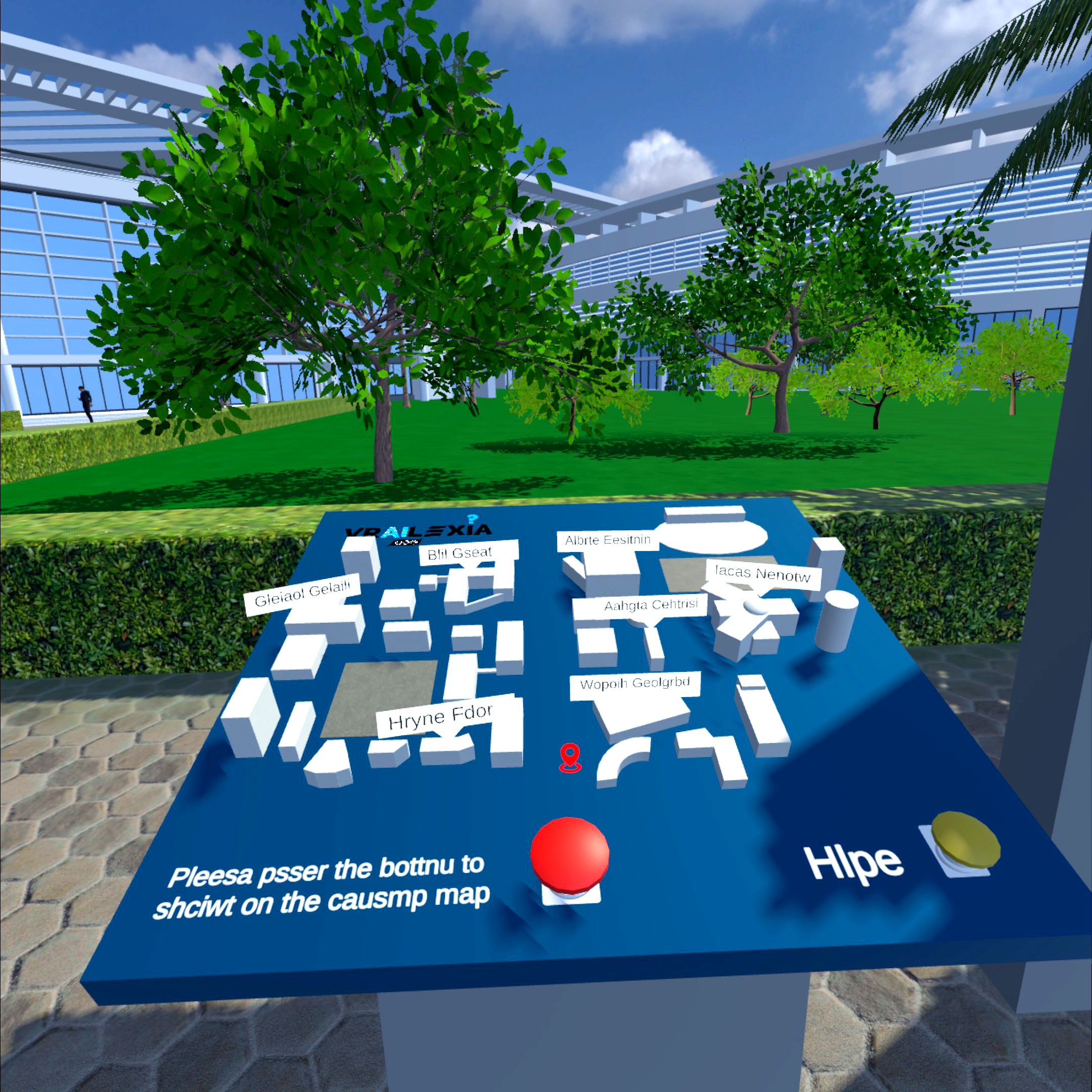}
                 \caption{\centering Interactive~3D Map.}
                 \label{fig:h_map}
            \end{subfigure}
            \caption{Maps.}
            \label{fig:maps}
        \end{figure}
    \item \textbf{{Students}}: The campus environment features diverse avatars representing students engaged in common activities, such as group interactions or solitary walks (see Figure~\ref{fig:elements}a). These non-playable characters (NPCs) add vibrancy and realism to the experience, simulating the bustling nature of a real university setting. However, participants cannot interact with these avatars, nor will the NPCs provide guidance or assistance, further reinforcing the sense of independence and potential disorientation.
    \item \textbf{{Signals}}: Informative signage is strategically placed throughout the campus to guide participants in completing challenges and navigating to key locations (see Figure~\ref{fig:elements}b). Initially, these signs provide clear instructions (tutorial); however, as~the experience progresses, some signs become increasingly difficult to read, simulating the textual decoding challenges often experienced by individuals with dyslexia.
In addition to directional signs, another set of signs identifies the entrances to campus buildings, displaying their respective names to assist with orientation. Some signs are dynamically modified based on the user's behavior. For~example, the~signs at the building entrances will indicate, when a participant attempts to enter the wrong building for the current task, that they are not in the correct building. This dynamic feature helps guide the user and reinforces the navigation challenges inherent to the experience.
This dual purpose, navigation and simulation, underscores the barriers associated with reading and spatial awareness.
    \item \textbf{{Beacon}}: In order to facilitate task progression and reduce uncertainty, a~visual beacon system is employed to highlight critical points within the campus. The~beacon appears as a circle of light projected onto the ground, signaling locations such as building entrances, exits, and~other significant waypoints (see Figure~\ref{fig:elements}c). At~any given time, the~active beacon guides participants sequentially through key stages of the experience. Additionally, there is a challenge beacon, observable from anywhere on the campus, that provides a clear indicator of the active 3D map corresponding to the current challenge. This helps users stay oriented and focused on the specific task at hand. The~structured approach of the beacon system aids participants in navigating simulated barriers and completing tasks more~effectively.

    \begin{figure}[H]
            
            \begin{subfigure}[b]{0.32\textwidth}
                 \centering
                 \includegraphics[width=\textwidth]{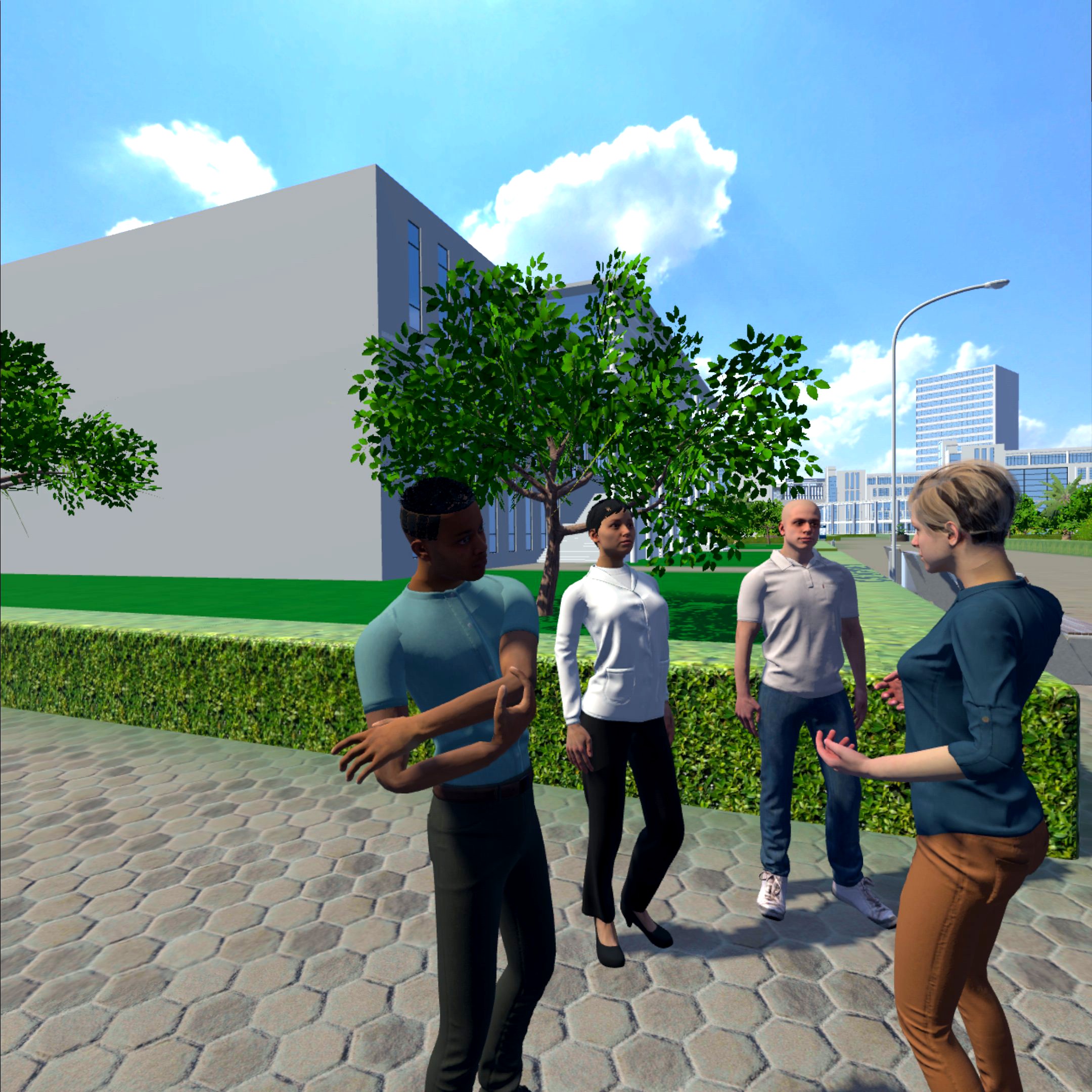}
                 \caption{\centering People}
                 \label{fig:people}
            \end{subfigure}
            \begin{subfigure}[b]{0.32\textwidth}
                 \centering
                 \includegraphics[width=\textwidth]{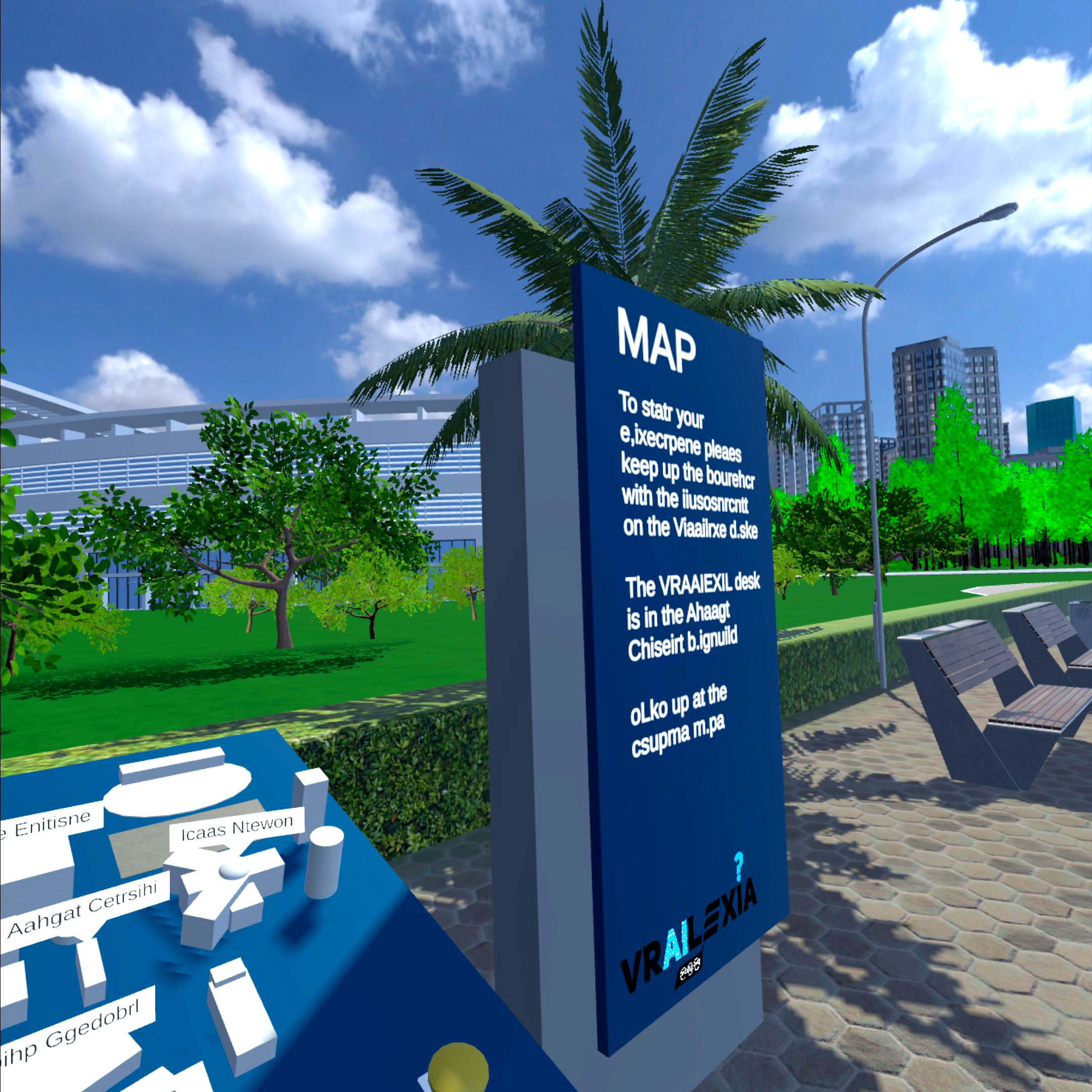}
                 \caption{\centering Signals}
                 \label{fig:signals}
            \end{subfigure}
            \begin{subfigure}[b]{0.32\textwidth}
                 \centering
                 \includegraphics[width=\textwidth]{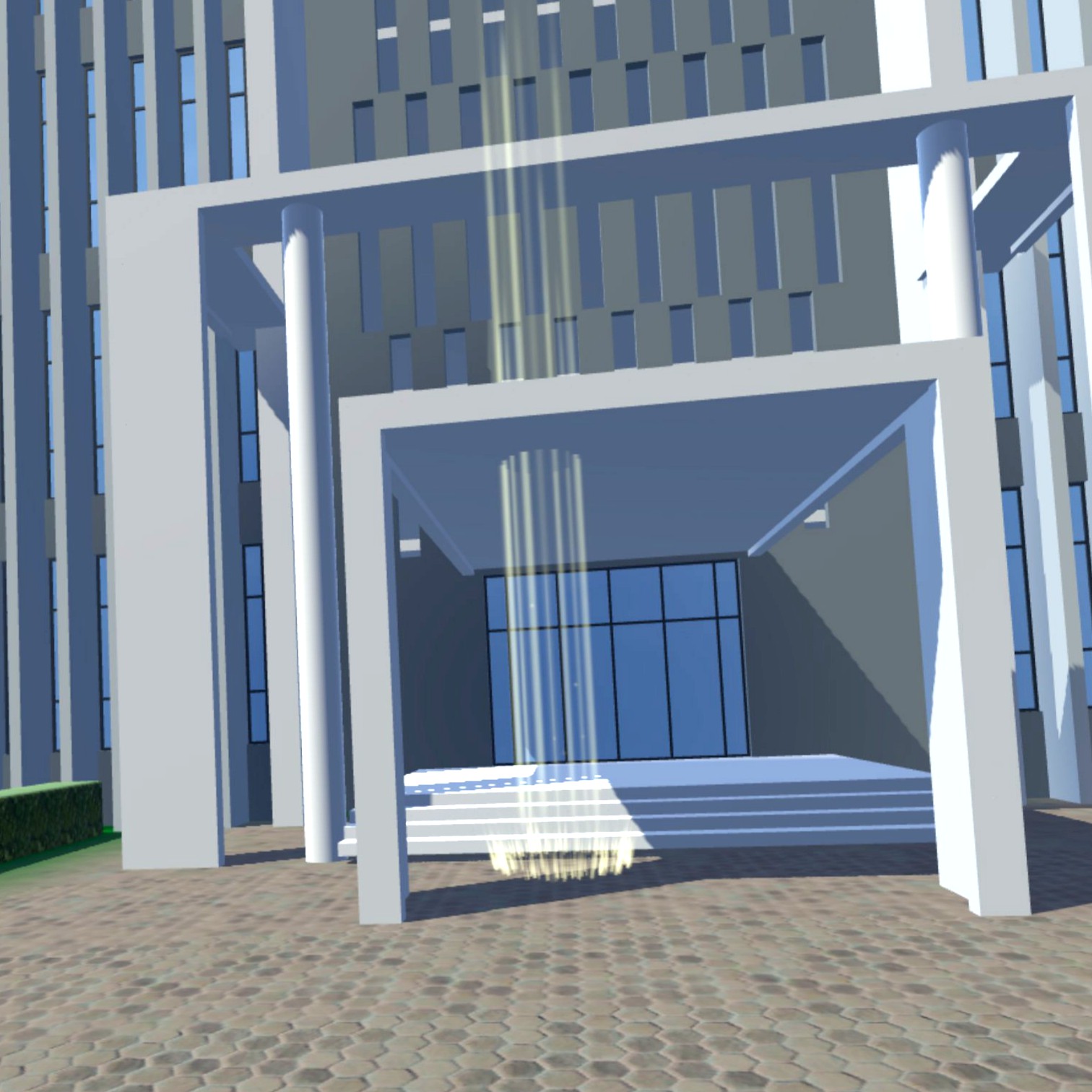}
                 \caption{\centering Beacon}
                 \label{fig:beacon}
            \end{subfigure}
            \vspace{3pt}
            \caption{Different elements included in the virtual~campus.}
            \label{fig:elements}
        \end{figure}
\end{itemize}

\subsection{Interaction}

\begin{figure}[H]
    
    \includegraphics[width=0.95\linewidth]{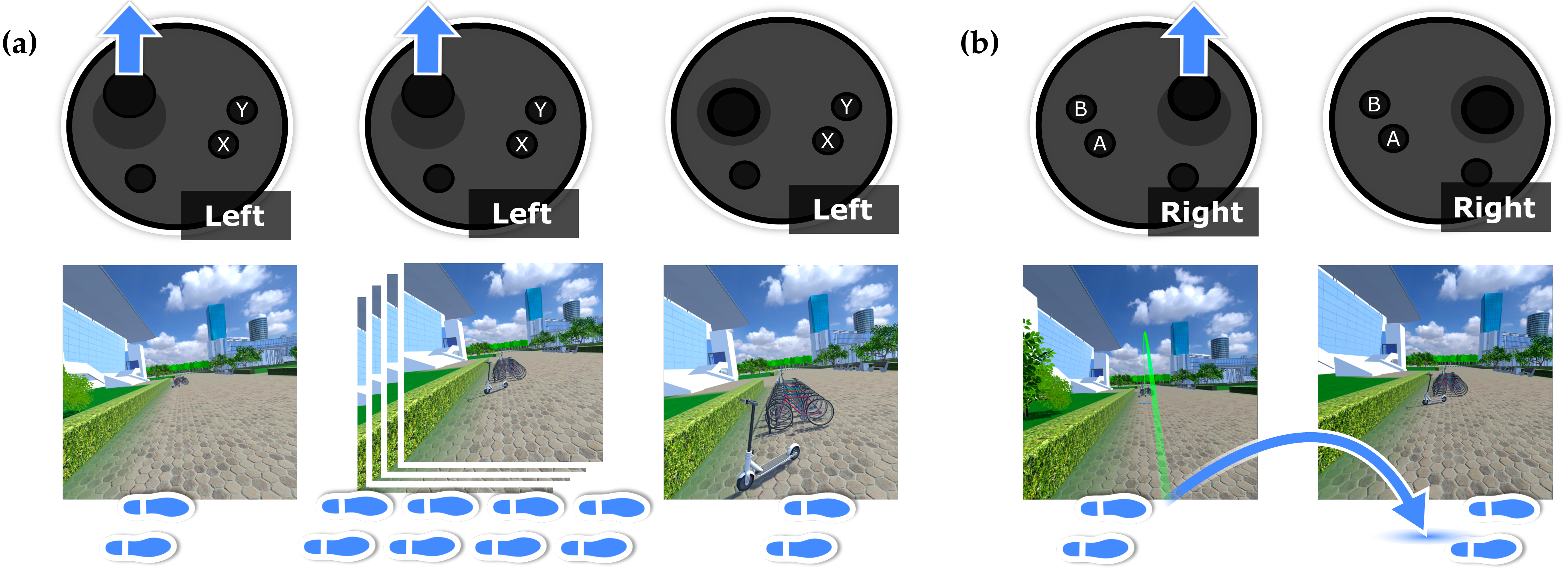}
    \caption{{Locomotion} modes using the controllers. {(\textbf{a})} Continuous locomotion. {(\textbf{b})} Teleportation.}
    \label{fig:locomotion}
\end{figure}

Within the virtual environment, users can engage in various types of interactions, each designed to enhance immersion and provide different ways of navigating and manipulating the virtual space. The~experience is implemented in 6 Degrees of Freedom (6DoF), allowing users to move freely along all three translational axes (forward/backward, up/down, left/right), as well as rotate along all three rotational axes (pitch, yaw, and~roll). This degree of freedom contributes significantly to a more immersive and realistic experience~\citep{DoFRossi2023}. The~interactions are facilitated through the use of controllers, which are essential for completing tasks and experiencing the environment fully. The~primary interaction types are as follows:

\begin{itemize}
    \item \textbf{{Continuous locomotion}}: This interaction simulates continuous movement through the virtual environment, allowing users to move freely along the three translational axes and rotate as they explore the space. Continuous locomotion offers precise and uninterrupted navigation, providing a high level of immersion. However, it can induce motion sickness due to the sensory mismatch between visual input and the vestibular system, and~in room-scale setups, it is limited by the physical space available. See Figure~\ref{fig:locomotion}a.

    \item \textbf{{Teleportation}}: Also known as parabolic locomotion, this interaction enables users to instantly jump to a selected point in the virtual environment. By~pointing at a destination, users are teleported through a parabolic arc, overcoming the physical space constraints. Teleportation provides an alternative to continuous locomotion, making it especially useful for large virtual environments. While it reduces motion sickness, it can disrupt immersion and hinder the creation of a mental map of the environment. However, this issue is minimized in open spaces like the campus, where the layout and open areas reduce the need for frequent teleportation and help maintain spatial orientation. See Figure~\ref{fig:locomotion}b.

    \item \textbf{{Button press}}: This interaction allows users to activate virtual buttons by bringing the controller close to the corresponding button in the environment, similar to physically pressing a button. This action involves rotational DoF (via wrist or arm movement), but~it may also require some translational movement to position the controller accurately. Button pressing is typically used for interacting with 3D map buttons or virtual controls, offering a simple and effective way to engage with the~environment.

    \item \textbf{{Grab/release objects}}: This interaction enables users to pick up objects within the virtual environment by positioning the controllers over them and pressing the grab button. Both translational and rotational DoF are required to manipulate objects effectively. Releasing an object is accomplished by simply releasing the corresponding button. This method is integral to tasks involving the handling and transportation of objects, enhancing the realism and tactile quality of the virtual experience.
\end{itemize}

These four interaction types, facilitated by 6 DoF, constitute the foundation of user engagement within the virtual environment. The~integration of translational and rotational movements, combined with diverse interaction modalities, results in a highly immersive and adaptable experience, thereby enhancing user presence and engagement in the simulated~environment.

\subsection{Simulating Barriers for~Dyslexics}
\label{sec:barriers}

The main objective of the experience is to raise participants' awareness of the barriers encountered by students with dyslexia during their higher education years. To~achieve this, a~series of simulated barriers have been designed and integrated into the virtual environment to reflect similar challenges. It is important to emphasize that these barriers are not intended to replicate dyslexia itself but to mimic the difficulties it can cause, such as slower reading, disorientation, and~feelings of isolation or lack of~support.

One of the primary challenges associated with dyslexia is difficulty in reading and comprehending text. To~simulate this challenge, two distinct strategies have been employed: letter movement and word~substitution.

\begin{figure}[H]

    \includegraphics[width=0.95\linewidth]{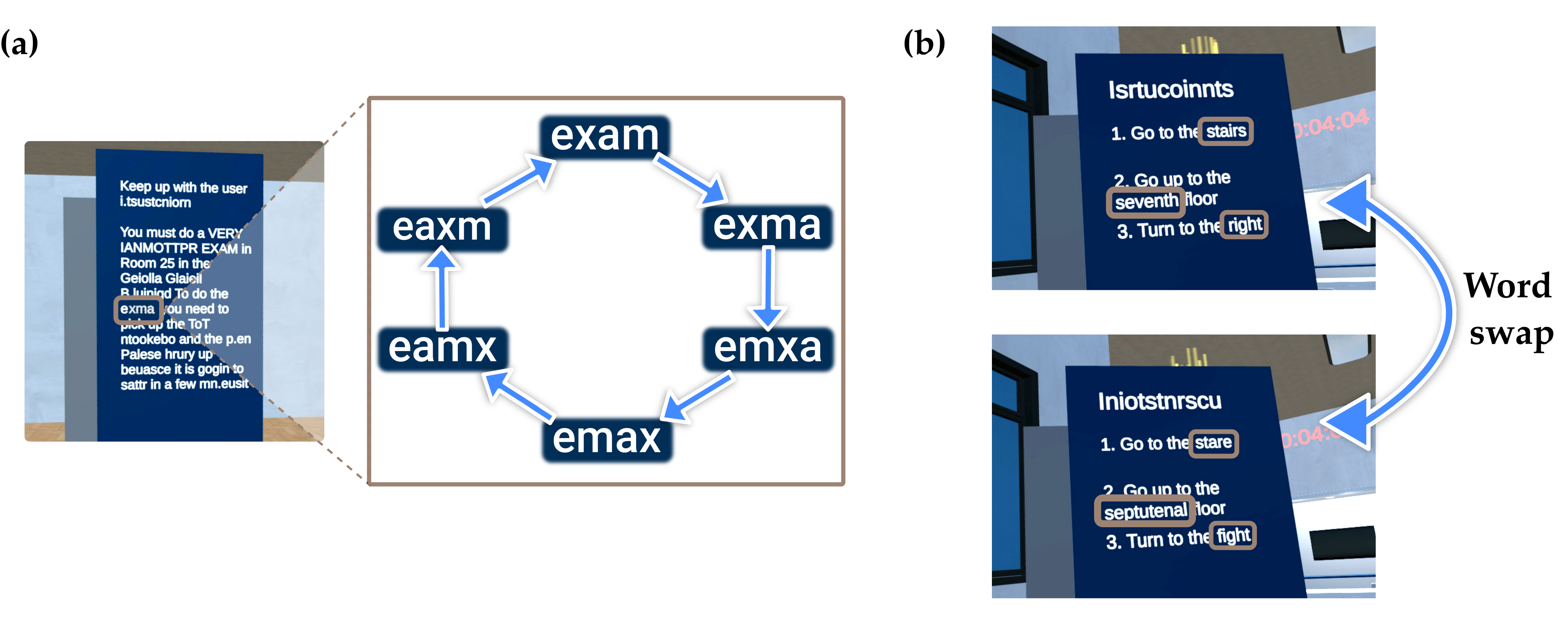}
    \caption{{Simulation} 
 of reading difficulties. {(\textbf{a})} Letter movement. {(\textbf{b})} Word~swapping.}
    \label{fig:reading_simulation}
\end{figure}

\begin{itemize}
    \item \textbf{{Letter movement:}} This barrier involves  dynamically and randomly rearranging the letters, except~for the first letter of the word, within~individual words, as~illustrated in Figure~\ref{fig:reading_simulation}a. In~this example, the~letters of the word ``exam'' change their order, forming nonexistent words like ``exma''. This effect is applied to the full text of informative signs that describe tasks on the campus and also in the names of the buildings that are reflected in the maps, making the text harder to read and increasing the time needed for comprehension.
    \item \textbf{{Word swapping:}} This barrier involves replacing some words in the text with other existing words that have similar sounds, which is one of the primary issues associated with surface dyslexia~\citep{lukov_dissociations_2015}. In~this way, during~the experience, words in task instructions are continuously swapped, forcing the user to pause and consider what the actual instructions are. This is illustrated in Figure~\ref{fig:reading_simulation}b, where the word ``stairs'' is replaced with ``stare''.
\end{itemize}

\begin{figure}[H]

    \includegraphics[width=0.9\linewidth]{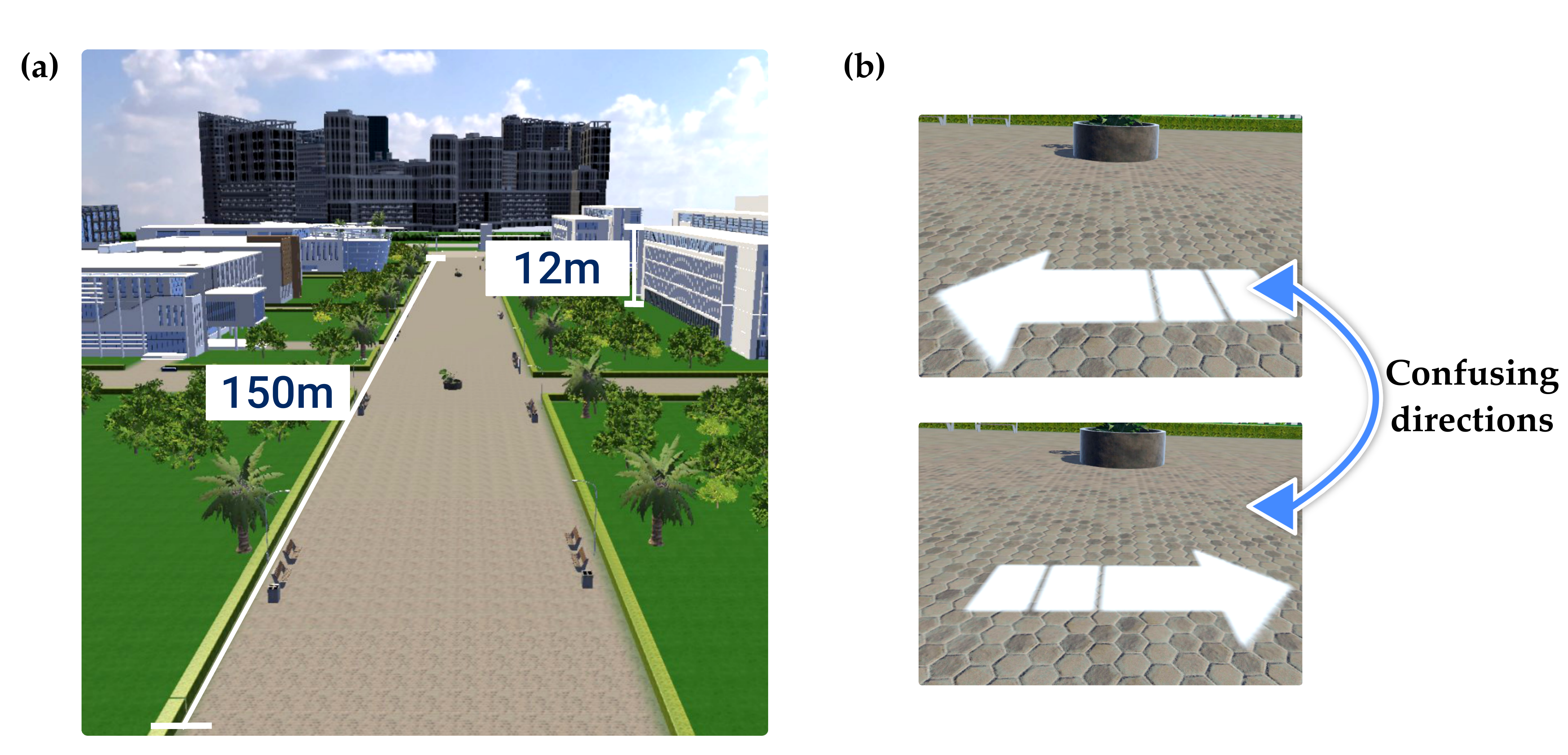}
    \caption{Simulation of orientation barriers. {(\textbf{a})} Scaled dimension of the virtual campus. {(\textbf{b})} Help option providing right and wrong~directions.}
    \label{fig:orientation}
\end{figure}

Another significant challenge faced by university students with dyslexia is orientation within large environments~\citep{disorientationCaldani2021}. These environments are often vast, with~similar-looking buildings and limited signage to guide navigation between locations. To~replicate this challenge, the~virtual campus was designed as a large-scale map featuring similar paths and structures. The~scaled dimensions of the buildings and streets within the modeled campus are depicted in Figure~\ref{fig:orientation}a. Furthermore, the~3D maps represent the shapes of buildings in a manner that does not fully correspond to their actual structures within the virtual environment. Similarly, vertical maps omit the names of the main buildings, further complicating navigation. This design compels users to rely on both types of maps to orient themselves effectively and reach their~destination.

Additionally, a~help button was incorporated into the 3D maps, enabling participants to display guiding arrows on the ground, as~illustrated in Figure~\ref{fig:orientation}b, to~assist them in navigating to their destination. However, to~ensure that this feature does not entirely mitigate the orientation challenge, the~arrows are deliberately designed to point in the wrong direction 25\% of the time, introducing an additional layer of disorientation. This design emulates some of the barriers encountered by individuals with dyslexia, particularly difficulties with left-right or up-down orientation. By~occasionally providing incorrect directions, this feature simulates the disorientation and navigational challenges that individuals with dyslexia may face in spatial~environments.

Another significant barrier faced by students with dyslexia is the lack of support from peers and teachers~\citep{Rohmer2022}. This sense of isolation and frustration often exacerbates the difficulties of their academic journey and can, in~some cases, result in the abandonment of their studies. To~replicate this experience, participants are placed in a campus environment populated with NPCs representing students and peers, but~interaction with them for assistance is not permitted. Furthermore, the~support provided through the virtual experience itself is intentionally limited and, at~times, deliberately misleading. This design choice seeks to heighten the user’s sense of frustration, transforming seemingly straightforward tasks into complex challenges and effectively simulating the emotional and practical difficulties experienced by students with~dyslexia.

\subsection{Gameplay}

The experience is designed to span multiple locations across the campus and is structured into distinct stages or levels. This section provides a detailed analysis of each phase of the experience, including the specific tasks participants are required to complete at each~stage.

Figure~\ref{fig:levels} illustrates the layout of these levels, mapping the progression from the user’s entry into the virtual environment to the conclusion of the experience. Each stage, along with the corresponding tasks, is classified into three main categories, which are outlined~below:

\begin{itemize} 
    \item \textbf{{Start--End Stages}
    }: These stages mark the beginning and conclusion of the experience. They include the introductory tutorial at the start and the final exam, where the experience comes to an end. 
    \item \textbf{{Main Building Task}}: These stages involve specific challenges within certain campus buildings, where participants face dyslexia-related barriers, particularly those related to reading and text comprehension. 
    \item \textbf{{Navigation Stage}}: These stages require participants to locate specific points on the campus to progress to the next phase. While the start and end points of these stages are predefined, as shown in Figure~\ref{fig:levels}, the~path taken by the player is not fixed and is influenced by the user’s navigation choices. 
\end{itemize}

The user progresses through a series of stages, each designed to represent a distinct type of interaction and challenge within the virtual campus. The~following breakdown categorizes the levels by their specific tasks and explains their relationship to the previously described phase types: Start--End Stages, Main Building Tasks, and~Navigation~Stages.

\begin{enumerate}
    \item \textbf{{Tutorial Stage (Start--End Stage):}}
    The experience begins with a tutorial stage located at the campus entrance. The~primary objective of this stage is to familiarize users with the controls and basic navigation mechanics of the virtual environment. Informative signs provide guidance on operating the controllers and selecting between different locomotion options to navigate to the first 3D interactive map. Upon~reaching the map, users receive instructions on interacting with elements such as virtual buttons. Successfully pressing the button transitions the experience into the next phase, marking the formal commencement of the user's task-based~journey.

    \item \textbf{{Journey 1 (Navigation Stage):}}
    Following the tutorial, the~user encounters their first navigation challenge. A~sign displaying instructions directs them to their next destination, utilizing the ``Letters Movement'' barrier to simulate the difficulties individuals with dyslexia may experience when interpreting text (see Section~\ref{sec:barriers}). The~instructions indicate that the user must proceed to the Agatha Christie building to collect materials required for an exam. This navigation task requires the user to interpret signs displaying building names and follow them to reach the designated location, thereby testing their ability to navigate the campus under conditions analogous to those faced by individuals with~dyslexia.

    \item \textbf{{Agatha Christie (Main Building Task):}}
    Upon entering the Agatha Christie building, the~user encounters another barrier: a sign displaying text altered by the ``Letters Movement'' effect. The~task requires the user to locate a notebook and pen within the building, collect them, and~place them in their virtual backpack before proceeding to the exam room. Upon~completing this task, the~user is informed that the exam room is located in the Galileo Galilei building. This challenge not only simulates the difficulties associated with reading impairments but also underscores the importance of time management, as~the user must complete the task within a limited time~frame.

    \item \textbf{{Journey 2 (Navigation Stage):}}
    After exiting the Agatha Christie building, the~user must navigate to the Galileo Galilei building using the newly provided instructions. This stage introduces a new 3D map to assist the user in locating the building. The~map includes an interactive help function, which offers additional guidance when required. The~user is tasked with traversing the campus to reach the Galileo Galilei building, marking a key milestone in this phase of the~experience.
\begin{figure}[H]
  
    \includegraphics[width=0.95\linewidth]{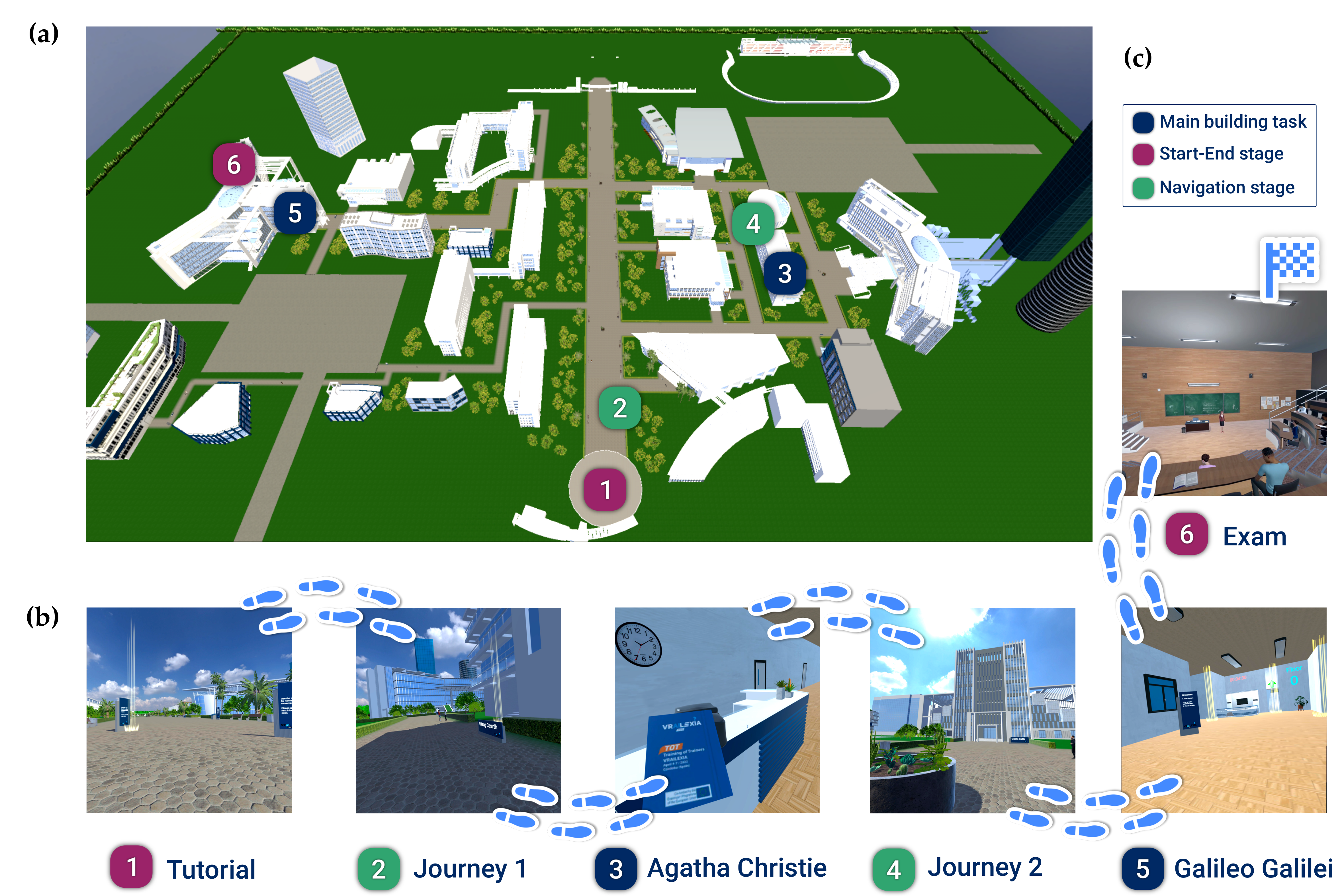}
    \caption{{(\textbf{a})} Campus render indicating the locations of the main stages of the experience. {(\textbf{b})} Flow of experience stages. {(\textbf{c})} Legend representing the different types of~levels.}
    \label{fig:levels}
\end{figure}
    \item \textbf{{Galileo Galilei (Main Building Task)}:}
    Upon entering the Galileo Galilei building, the~user encounters a room featuring a staircase, a~countdown timer, and~two hallways. A~new sign appears, utilizing the ``Swapping Words'' barrier to simulate another challenge faced by individuals with dyslexia---difficulty with word recognition. This sign provides instructions specifying the floor and hallway the user must follow to locate the exam room. The~countdown timer introduces a time-sensitive element, replicating the stress and pressure that students may experience during~exams.

    \item \textbf{{Exam Classroom (Start--End Stage)}:}
    Finally, the~user reaches the exam room, where they receive instructions to begin the exam. These instructions, while seemingly straightforward, contain ``swapped'' words that resemble real words but are, in~fact, nonsensical (e.g., ``Go to your zona'' instead of ``Go to your zone''). This final task simulates the challenges of reading comprehension and word recognition challenges commonly experienced by individuals with dyslexia in academic settings. Upon~completing this task, the~experience concludes.
\end{enumerate}

An example of the gameplay is available on the YouTube platform ( 
 \url{https://youtu.be/j0hBD1j5bgc}, 
 {accessed on 19 February 2025}). 
  The~complete experience is available as a free download via SideQuest ( \url{https://sidequestvr.com/app/37775}, {accessed on 19 February 2025}). 


\section{Evaluation of the~Experience}\label{sec4}

This section details the methodological framework used to evaluate the effectiveness of the proposed VR experience in fostering empathy and raising awareness about dyslexia. The~assessment is structured into three key aspects: the features of the participants involved in the study, the~methods employed for data collection and analysis, and~the instruments used to measure the impact of the~experience.

\subsection{Participants}
To assess the effectiveness of the proposed method as a VR experience aimed at raising awareness about dyslexia, an~initial evaluation was conducted with 32 participants from the higher education sector. These participants were selected from volunteers who expressed interest in the experience during a project's dissemination campaign carried out at the University of Córdoba (Spain). Socio-demographic data were collected during the participant selection process to analyze the methodology's applicability across different demographic groups, with~a primary focus on gender, age, and participants' academic or professional roles within the higher education context. Table~\ref{tab:participant_features} contains information on these characteristics for the 32 participants selected for this study. During~the selection process, efforts were made to maintain a balance in the gender of participants, ultimately resulting in a slightly higher participation of male participants, comprising approximately 59\% of the total sample. Regarding profiles, the~majority of the experience was conducted with university students who shared classes with dyslexic students, representing the largest group (53.17\%). Nonetheless, a~significant number of teachers of students with dyslexia also showed interest in the experience (21.88\%), as~well as other individuals who, despite not being educationally connected to someone with dyslexia, had a relative with dyslexia. Concerning age, as~previously mentioned, the~tool was primarily used by students, making the predominant age range between 18 and 26 years, with~over 60\% of participants falling within this~range.

\begin{table}[H]
\caption{Socio-demographic features of the participants in the~study.\label{tab:participant_features}}
\footnotesize
\begin{adjustwidth}{-\extralength}{0cm}
    \newcolumntype{C}{>{\centering\arraybackslash}X}
    \begin{tabularx}{\fulllength}{CcCC}
        \toprule
        \textbf{Feature} & \textbf{Subgroup} & \textbf{\textit{\#}} & \textbf{Percentage} \\
        \midrule
        \multirow[m]{3}{*}{Gender}
            & Female               & 13 & 59.38 \\
            & Male                 & 19 & 40.62 \\
            & Other                & 0  & 0.00 \\
        \midrule
        \multirow[m]{3}{*}{Age} 
            & [18, 26)             & 20 & 62.5 \\
            & [26, 50]             & 10 & 31.25 \\
            & >50                 & 2 & 6.25 \\
        \midrule
        \multirow[m]{4}{*}{Participant profile}
            & Dyslexic relative    & 6 & 18.75 \\
            & Higher education student   & 17 & 53.12 \\
            & Higher education professor & 7 & 21.88 \\
            & None                & 2 & 6.25 \\
        \bottomrule
    \end{tabularx}
\end{adjustwidth}
\end{table}
\unskip

\subsection{Methods}
The evaluation was conducted through a controlled study with a structured data collection process, ensuring the validity and reliability of the findings. The~primary aim was to determine how the VR experience influences users' understanding of dyslexia-related challenges. In~order to do that a selected group of participants engaged in the VR experience under supervised conditions, after~which they were asked to complete a structured questionnaire that will be defined~later.

The theoretical framework guiding the immersive learning experience is based on constructivist and experiential learning theories. Constructivist and experiential learning theories emphasize active, learner-centered engagement, where participants acquire knowledge through meaningful interactions with their environment. In~this context, the~VR experience was designed to provide an immersive simulation that exposes users to the barriers faced by dyslexic individuals, encouraging a deeper understanding through direct experience and reflection~\cite{Lie2023}.
Additionally, the~cognitive theory of multimedia learning informed the integration of visual and interactive elements to enhance engagement and retention. By~simulating real-world scenarios and embedding dyslexia-related challenges, the~experience aligns with these pedagogical principles, ensuring that learners actively engage with the presented difficulties, rather than passively receiving~information.

All participants provided informed consent before engaging in the study, ensuring they were fully aware of the research objectives and the handling of their data. The~study complied with all ethical guidelines, including Organic Law 3/2018 on Personal Data Protection and the ethical principles outlined in the Declaration of Helsinki~\citep{LOPDGDD2018,helsinki20123}. Ethical approval for the study was granted by the institutional ethics committee (approval number 367). In~addition, to~ensure data privacy, all information obtained was treated with strict confidentiality. The~questionnaire was designed to be anonymous, including an introductory section explaining the study's objectives, followed by an informed consent~form.

\subsection{Instruments}
To evaluate the quality of the experience and its effectiveness as a tool for raising awareness about dyslexia, a~questionnaire-based evaluation methodology was developed. The~questionnaire consists of 15 questions to be completed after participating in the experience, divided into three categories: socio-demographic data, VR experience quality, and~dyslexia awareness. The~design of the questions was informed by recommendation models aimed at supporting dyslexic students~\citep{Morciano2024} combined with the VR dimension of the experience. Finally, a~general evaluation question regarding the overall experience is also included. Table~\ref{tab:survey} presents the list of questions organized by category. All items are answered using a 5-point Likert scale except for socio-demographic~questions.

The collected data provided quantitative insights into how users perceived the VR experience and whether it effectively enhanced their understanding of dyslexia-related challenges. The~next section presents and discusses the results obtained from the~evaluation.
\begin{table}[H]
\caption{Survey questions divided by~categories.\label{tab:survey}}
\begin{adjustwidth}{-\extralength}{0cm}
\newcolumntype{K}{>{\centering\arraybackslash}X}
\begin{tabularx}{\fulllength}{KK}
\toprule
\textbf{Category} & \textbf{Question} \\ 
\midrule
\multirow[m]{3}{*}{Socio-demographic information} & Gender \\ 
                                            & Age\\ 
                                            & Relationship to a dyslexic student *\\ 
\midrule
\multirow[m]{4}{*}{VR experience quality}   & VR prior~experience \\
                                            & Motion~sickness \\
                                            & Visual~immersion \\
                                            & Interaction within the~environment \\
                                            \bottomrule
\end{tabularx}
\end{adjustwidth}
\end{table}
\begin{table}[H]\ContinuedFloat
\caption{\textit{Cont.}\label{tab:survey}}
\begin{adjustwidth}{-\extralength}{0cm}
\newcolumntype{K}{>{\centering\arraybackslash}X}
\begin{tabularx}{\fulllength}{KK}
\toprule
\textbf{Category} & \textbf{Question} \\ 

\midrule
\multirow[m]{3}{*}{Barrier simulation}     & Navigation among~buildings \\
                                            & Finding the exam~class \\
                                            & Use of the help~option \\
\midrule
\multirow[m]{3}{*}{Dyslexia awareness}      & Awareness of~dyslexia \\
                                            & Experience with empathy~applications \\
                                            & Effectiveness of the virtual campus in promoting dyslexia~awareness \\
\midrule
Overall                                     & Overall rating of the~experience \\
\bottomrule
\end{tabularx}
\end{adjustwidth}
\noindent{\footnotesize{* The following types of relationships with a dyslexic student were considered: (1) classmate (2) teacher (3) relative (4) other (5) none. }}
\end{table}

\section{Results and~Discussion}\label{sec5}

This section presents and discusses the results of the study, focusing on the effectiveness of the empathy experience both as a VR application and as a tool for raising awareness about dyslexia. The~results are structured into two main aspects: the assessment of the VR application in terms of usability and immersion, and~the evaluation of its effectiveness in simulating barriers and raising awareness about~dyslexia.

\subsection{VR~Application}
\begin{figure}[H]
    \centering
    \includegraphics[width=0.98\linewidth]{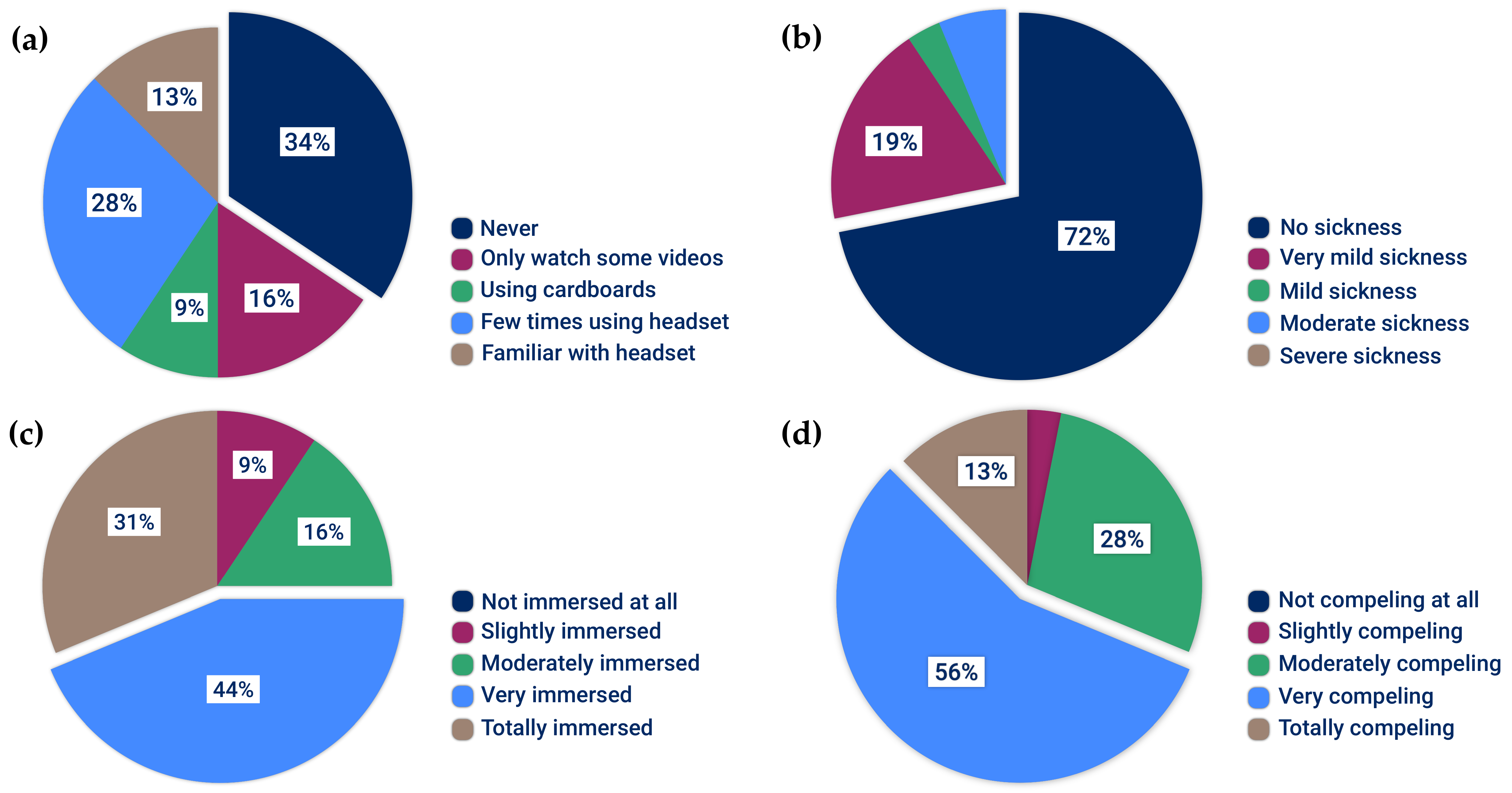}
   \caption{{Responses} 
 to the questions measuring the quality of the VR application. {(\textbf{a})} Have you ever tried VR before the experience? {(\textbf{b})} Did you experience nausea, loss of balance, or~discomfort during the experience? {(\textbf{c})} How much did the visual aspects of the environment engage you? {(\textbf{d})}~How compelling was your sense of moving around and interacting within the virtual environment?}
    \label{fig:pie_vr}
\end{figure}

In order to assess the quality of the proposed tool, it is critical to evaluate not only its effectiveness in achieving its primary objective, raising awareness about dyslexia, but~also its ability to effectively utilize the unique features of VR. Therefore, the~first evaluation aspect focused on assessing the quality of the VR experience in terms of immersion, interaction, and~user comfort. Figure~\ref{fig:pie_vr} illustrates the responses to four key questions regarding prior VR experience, motion sickness, visual immersion, and~interaction within the virtual~environment.

The questionnaire results revealed that a significant proportion of participants had no prior experience with VR, making this their first encounter with a complete setup involving a headset and controllers. Conversely, 28\% of participants reported familiarity with VR equipment, providing valuable feedback based on their previous exposure to other VR applications or games.
Regarding discomfort or motion sickness, the~findings were predominantly positive. The~majority of participants reported no sickness, with~only one individual experiencing moderate symptoms and none reporting severe discomfort.
In terms of immersion, participants generally expressed favorable opinions. Specifically, 44\% indicated feeling ``Very immersed'', and~16\% reported being ``Totally immersed'' in the virtual environment. Among~the remaining participants, the~most commonly cited issue was that the use of controllers for locomotion, particularly teleportation, diminished their sense of immersion.
Finally, most participants expressed high satisfaction with the interaction mechanics and locomotion options within the virtual environment. However, the~primary criticism is centered on the limited physical space available, which restricted freer physical~movement.

\subsection{Barrier Simulation and Dyslexia~Awareness}

The objective of this study is to raise awareness about the challenges faced by dyslexic students in higher education. To~evaluate the effectiveness of the proposed tool in achieving this goal, the~evaluation questions were divided into two categories: the simulation of barriers and awareness of dyslexia. The~first category focused on assessing the extent to which the simulated barriers hinder task completion during the VR experience. The~second category aimed to evaluate participants' prior knowledge of dyslexia as an SLD and how their perceptions of dyslexic students are influenced after participating in the~experience.

\begin{figure}[H]
    
    \includegraphics[width=0.98\linewidth]{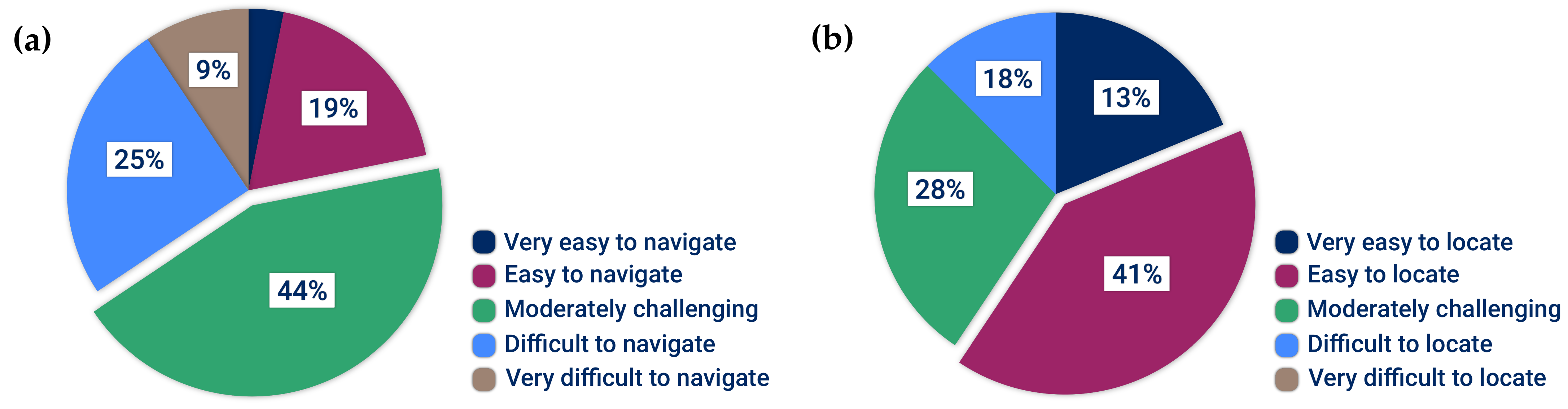}
    \caption{{Responses} 
 to the questions measuring the effectiveness of the application in simulating barriers. {(\textbf{a})} How challenging did you find navigating among the buildings? {(\textbf{b})} How challenging did you find locating the classroom for the exam?}
    \label{fig:pie_barriers}
\end{figure}

Regarding the barriers, Figure~\ref{fig:pie_barriers} illustrates the responses obtained when participants were asked about the difficulties encountered while performing the two main tasks of the~experience. 

The first task required participants to navigate between various campus buildings (see Figure~\ref{fig:pie_barriers}a). During~this task, participants encountered barriers related to orientation within a monotonous and large-scale environment, as~well as difficulties in reading the instructions on signs due to the implemented ``Letter movement'' strategy. To~address these challenges, the~experience included mitigation measures, such as the availability of various maps and a help option that displayed guiding arrows on the ground to guide participants between buildings. Despite these aids, only 12 out of 32 participants utilized the help function, as~most preferred to complete the task without additional assistance. Participants generally identified this task as the most challenging, citing their main difficulty as transferring the information from the provided maps, particularly the 3D interactive maps, to~the virtual campus~space.

The second task involved finding the exam classroom within the building where the final exam takes place (see Figure~\ref{fig:pie_barriers}b). Upon~entering the building, participants are presented with confusing instructions generated using the ``Word swapping'' strategy. This challenge is further intensified by the monotonous design of the building's floors, which makes it difficult for participants to easily distinguish between them. Despite these barriers, participants generally found this task to be straightforward, noting that the allotted time was more than sufficient. While some participants needed to review the instructions multiple times to fully understand them, the~overall simplicity and limited number of steps contributed to the task being perceived as~easy.

\begin{figure}[H]
    
    \includegraphics[width=0.98\linewidth]{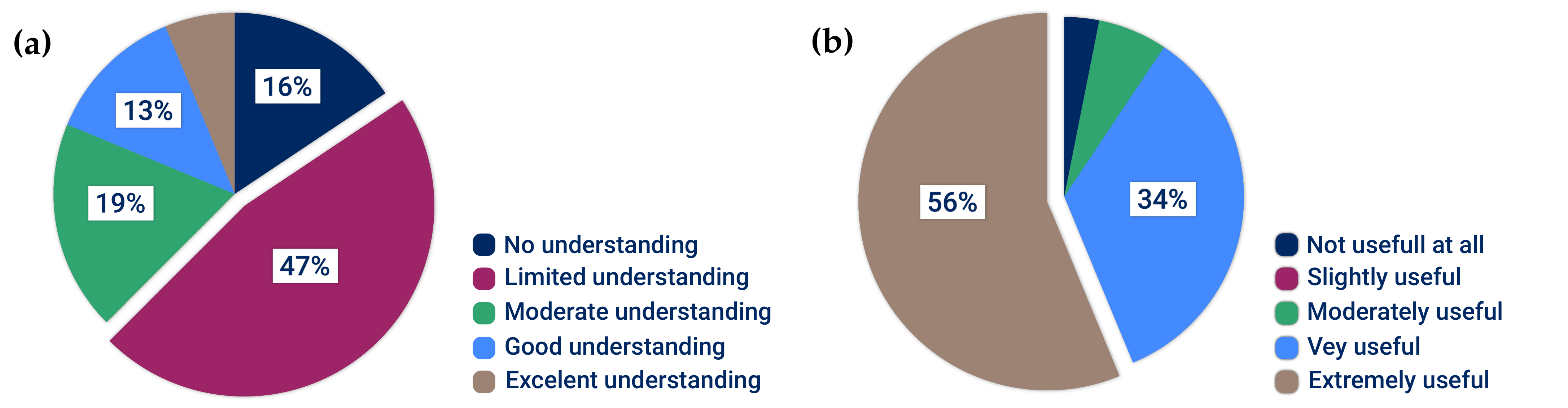}
    \caption{{Responses} 
 to the questions about participants' dyslexia awareness. {(\textbf{a})} How would you rate your understanding of dyslexia? {(\textbf{b})} How useful do you consider the VR experience in raising 
  awareness about dyslexia?}
    \label{fig:pie_awar}
\end{figure}

On the other hand, the~responses to the questions regarding the awareness of dyslexia are shown in Figure~\ref{fig:pie_awar}. 

The first notable finding is the general lack of knowledge about dyslexia, with~more than 60\% of participants either being unaware of what this learning disorder is or having only heard about it (see Figure~\ref{fig:pie_awar}a). In~contrast, more than 90\% of participants indicated that the campus experience is ``Very useful'' or ``Extremely useful'' in raising awareness about dyslexia (see Figure~\ref{fig:pie_awar}b). Participants frequently noted that they had not previously considered how the challenges associated with this learning disorder could result in significant difficulties when performing seemingly simple tasks. Only one participant reported finding the tool not useful for raising awareness about dyslexia. This individual stated that he did not fully understand the connection between the tasks and challenges presented during the experience or the definition of dyslexia~difficulties.

\subsection{Discussion}

The findings of this study highlight the potential of VR as an effective tool for raising awareness about dyslexia by immersing participants in the challenges faced by dyslexic students. These results align with previous research demonstrating that VR can enhance empathy for individuals with disabilities~\citep{Tong2020}. Compared to traditional awareness methods such as text-based or video interventions, the~interactive and embodied nature of VR provides a more engaging and impressive experience~\citep{Bryant2019}.

Beyond its practical contributions, the~study supports key theoretical frameworks. The~results align with embodied cognition theory, which suggests that active engagement enhances learning and understanding~\citep{Lin2024}. 
Situating these findings within the broader research landscape, this study contributes to the emerging field of VR-based disability awareness~\cite{hoter_effects_2022,Jin2023}. While VR has been extensively explored for cognitive training, including in the dyslexia field~\citep{Rello2016}, its application in fostering social awareness remains underdeveloped. The~results presented suggest that VR can effectively complement existing educational interventions by providing an experiential understanding of~dyslexia. 

Despite its promising results, the~study faced limitations. The~sample size was relatively small, and it focused on higher education participants. Additionally, another important limitation of the proposal was that, despite the software being freely available to users, a~Meta headset was required to participate in the experience. This requirement represents a significant barrier, as~the cost of such headsets may make them inaccessible to many potential~users.

In summary, the~study confirms that VR is a promising tool for dyslexia awareness, offering an immersive and engaging alternative to conventional methods. To~the best of our knowledge, no previous study has explored the use of VR to foster empathy specifically for dyslexic individuals, making this work a novel contribution to the field. By~situating the findings within prior research and theoretical frameworks, this work provides a foundation for future exploration of VR in inclusive education. The further refinement and expansion of these experiences will be essential to maximize their impact and~accessibility.

\section{Conclusions and Future~Work}\label{sec6}

The findings of this study offer valuable insights into the effectiveness of the proposed VR experience in simulating the barriers encountered by dyslexic students and raising awareness about this SLD among participants. Overall, the~tool demonstrated significant potential in achieving its awareness objectives, as~reflected by a high average participant rating of 4.72 out of 5. This high level of satisfaction indicates that the VR experience effectively engaged participants and provided a meaningful understanding of the challenges associated with~dyslexia. 

As future work, and~building upon the results of this initial evaluation, the~next phase of research, currently in progress, focuses on the development and implementation of a structured set of activities to evaluate the effectiveness of VR as a tool for promoting empathy and raising awareness about dyslexia among undergraduate healthcare students. Among~these activities, an~adapted version of the Toronto Empathy Questionnaire (TEQ)~\citep{Spreng2009} will be developed and contextualized for higher education students with dyslexia. This questionnaire will be administered both at the initial phase and post-intervention to check the impact of the~experience.

The study will consist of three phases: baseline assessment, intervention, and~follow-up evaluation. In~the baseline phase, students’ initial knowledge, empathy, and~awareness of dyslexia will be assessed through pre-intervention surveys. During~the intervention phase, third- and fourth-year healthcare students will participate in the VR-based program simulating the experiences of individuals with dyslexia. The participants will be randomly assigned to either an experimental group, which will engage with the VR intervention, or~a control group, which will receive conventional educational instruction. The~VR intervention comprises a 20-min immersive session illustrating the cognitive and emotional challenges associated with dyslexia.
In the post-intervention phase, a~follow-up survey will be conducted three weeks later~\citep{PrezLpez2024} to evaluate changes in empathy and awareness. The~customized TEQ outcomes will be compared between the experimental and control groups. The~findings of this study aim to refine educational strategies in healthcare training, ultimately fostering greater empathy and a deeper understanding of learning~disorders.



\vspace{6pt} 




\authorcontributions{
Conceptualization, J.M.A.-L., E.Y.-B., P.A.-M. and~S.P.; methodology, J.M.A.-L., E.Y.-B. and~P.A.-M.; software, J.M.A.-L. and E.Y.-B.; validation, J.M.A.-L., E.Y.-B., A.Z., P.A.-M. and~S.P.; formal analysis, J.M.A.-L.; investigation, J.M.A.-L., E.Y.-B. and~P.A.-M.; resources, E.Y.-B. and~S.P.; data curation, J.M.A.-L., E.Y.-B., A.Z. and~P.A.-M.; writing---original draft preparation, J.M.A.-L.; writing---review and editing, J.M.A.-L., E.Y.-B., A.Z., P.A.-M. and~S.P.; visualization, J.M.A.-L. and P.A.-M.; supervision, E.Y.-B., A.Z. and~S.P.; project administration, E.Y.-B. and S.P.; funding acquisition, S.P. All authors have read and agreed to the published version of the~manuscript.

}

\funding{{This} 
research was part of 
	``VRAIlexia –
Partnering Outside the Box: Digital and Artificial Intelligence
Integrated Tools to Support Higher Education Students with
Dyslexia''  funded by Erasmus + Programme 2014--2020 grant number 2020-1-IT02-KA203-080006.

}

\informedconsent{
Informed consent was obtained from all subjects involved in the~study.

}

\dataavailability{
The original contributions presented in this study are included in the
article. Further inquiries can be directed to the corresponding authors. 
} 




\acknowledgments{
José Manuel Alcalde Llergo is a {PhD student} 
 enrolled in the National PhD in Artificial Intelligence, XXXVIII cycle, course on health and life sciences organized by Università Campus Bio-Medico di Roma. He is also pursuing his doctorate under co-supervision at the Universidad de Córdoba (Spain), enrolled in its PhD program in Computation, Energy and {Plasmas}.
}

\conflictsofinterest{The authors declare no conflicts of interest.
} 






\begin{adjustwidth}{-\extralength}{0cm}

\reftitle{References}

\PublishersNote{}
\end{adjustwidth}

\begin{thebibliography}{999}

\bibitem[Xie et~al.(2021)Xie, Liu, Alghofaili, Zhang, Jiang, Lobo, Li, Li,
  Huang, Akdere, Mousas, and Yu]{Xie2021}
Xie, B.; Liu, H.; Alghofaili, R.; Zhang, Y.; Jiang, Y.; Lobo, F.D.; Li, C.; Li,
  W.; Huang, H.; Akdere, M.;  et~al.
\newblock A Review on Virtual Reality Skill Training Applications.
\newblock {\em Front. Virtual Real.} {\bf 2021}, {\em 2}, {645153.} 
\newblock {\url{https://doi.org/10.3389/frvir.2021.645153}}.

\bibitem[Zingoni et~al.(2021)Zingoni, Taborri, Panetti, Bonechi,
  Aparicio-Martínez, Pinzi, and Calabrò]{zingoni_investigating_2021}
Zingoni, A.; Taborri, J.; Panetti, V.; Bonechi, S.; Aparicio-Martínez, P.;
  Pinzi, S.; Calabrò, G.
\newblock Investigating {Issues} and {Needs} of {Dyslexic} {Students} at
  {University}: {Proof} of {Concept} of an {Artificial} {Intelligence} and
  {Virtual} {Reality}-{Based} {Supporting} {Platform} and {Preliminary}
  {Results}.
\newblock {\em Appl. Sci.} {\bf 2021}, {\em 11},~4624.
\newblock {\url{https://doi.org/10.3390/app11104624}}.

\bibitem[Benedetti et~al.(2022)Benedetti, Barone, Panetti, Taborri, Urbani,
  Zingoni, and Calabrò]{Benedetti2022}
Benedetti, I.; Barone, M.; Panetti, V.; Taborri, J.; Urbani, T.; Zingoni, A.;
  Calabrò, G.
\newblock Clustering analysis of factors affecting academic career of
  university students with dyslexia in Italy.
\newblock {\em Sci. Rep.} {\bf 2022}, {\em 12}, {9010}.
\newblock {\url{https://doi.org/10.1038/s41598-022-12985-w}}.

\bibitem[Butterfuss and Kendeou(2018)]{Butterfuss2018}
Butterfuss, R.; Kendeou, P.
\newblock The Role of Executive Functions in Reading Comprehension.
\newblock {\em Educ. Psychol. Rev.} {\bf 2018}, {\em 30},~1--26.
\newblock {\url{https://doi.org/10.1007/s10648-017-9422-6}}.

\bibitem[Farah et~al.(2021)Farah, Ionta, and Horowitz-Kraus]{Farah2021}
Farah, R.; Ionta, S.; Horowitz-Kraus, T.
\newblock Neuro-Behavioral Correlates of Executive Dysfunctions in Dyslexia
  Over Development From Childhood to Adulthood.
\newblock {\em Front. Psychol.} {\bf 2021}, {\em 12}.
\newblock {\url{https://doi.org/10.3389/fpsyg.2021.708863}}.

\bibitem[Miles et~al.(2013)Miles, Gilroy, and Du~Pre]{miles2013dyslexia}
Miles, T.; Gilroy, D.; Du~Pre, E.
\newblock {\em Dyslexia at College}; Taylor \& Francis:  {Abingdon, UK,} 
  2013.

\bibitem[Franceschini et~al.(2022)Franceschini, Bertoni, Puccio, Gori, Termine,
  and Facoetti]{Franceschini2022}
Franceschini, S.; Bertoni, S.; Puccio, G.; Gori, S.; Termine, C.; Facoetti, A.
\newblock Visuo-spatial attention deficit in children with reading
  difficulties.
\newblock {\em Sci. Rep.} {\bf 2022}, {\em 12}, {13930}.
\newblock {\url{https://doi.org/10.1038/s41598-022-16646-w}}.

\bibitem[Costantini et~al.(2020)Costantini, Ceschi, and
  Sartori]{costantini_psychosocial_2020}
Costantini, A.; Ceschi, A.; Sartori, R.
\newblock Psychosocial {Interventions} for the {Enhancement} of {Psychological}
  {Resources} among {Dyslexic} {Adults}: {A} {Systematic} {Review}.
\newblock {\em Sustainability} {\bf 2020}, {\emph{12}, 7994.} 
\newblock {\url{https://doi.org/10.3390/su12197994}}.

\bibitem[Association(2024)]{dyslexia_basics}
Association, I.D.
\newblock Dyslexia Basics. 2024.
\newblock Available online: \url{https://dyslexiaida.org/dyslexia-basics/} (accessed on 14 November 2024).


\bibitem[Snowling and Hulme(2024)]{Snowling2024}
Snowling, M.; Hulme, C.
\newblock Do we really need a new definition of dyslexia? A commentary.
\newblock {\em Ann. Dyslexia} {\bf 2024}, {\em 74},~355–362.
\newblock {\url{https://doi.org/10.1007/s11881-024-00305-y}}.

\bibitem[Lampropoulos and Kinshuk(2024)]{vrEducationReview}
Lampropoulos, G.; Kinshuk.
\newblock Virtual reality and gamification in education: a systematic review.
\newblock {\em Educ. Technol. Res. Dev.} {\bf 2024},
  {\em 72},~1691--1785.
\newblock {\url{https://doi.org/10.1007/s11423-024-10351-3}}.

\bibitem[Materazzini et~al.(2024)Materazzini, Melis, Zingoni, Baldacci,
  Calabrò, and Taborri]{Materazzini2024}
Materazzini, M.; Melis, A.; Zingoni, A.; Baldacci, D.; Calabrò, G.; Taborri,
  J.
\newblock Which Are the Needs of People with Learning Disorders for Inclusive
  Museums? Design of OLOS{\textregistered}—An Innovative Audio-Visual
  Technology.
\newblock {\em Appl. Sci.} {\bf 2024}, {\em 14},~3711.
\newblock {\url{https://doi.org/10.3390/app14093711}}.

\bibitem[Rohmer et~al.(2022)Rohmer, Doignon-Camus, Audusseau, Trautmann,
  Chaillou, and Popa-Roch]{Rohmer2022}
Rohmer, O.; Doignon-Camus, N.; Audusseau, J.; Trautmann, S.; Chaillou, A.C.;
  Popa-Roch, M.
\newblock Removing the academic framing in student evaluations improves
  achievement in children with dyslexia: The mediating role of self-judgement
  of competence.
\newblock {\em Dyslexia} {\bf 2022}, {\em 28},~309--324.
\newblock {\url{https://doi.org/10.1002/dys.1713}}.

\bibitem[Neugebauer et~al.(2024)Neugebauer, Castner, Severitt, Stingl, Ivanov,
  and Wahl]{vrVisualImapirmentSimulation}
Neugebauer, A.; Castner, N.; Severitt, B.; Stingl, K.; Ivanov, I.; Wahl, S.
\newblock Simulating vision impairment in virtual reality: a comparison of
  visual task performance with real and simulated tunnel vision.
\newblock {\em Virtual Real.} {\bf 2024}, {\em 28}, {97.}
\newblock {\url{https://doi.org/10.1007/s10055-024-00987-0}}.

\bibitem[Passig(2011)]{Passig2011}
Passig, D.
\newblock The Impact of Immersive Virtual Reality on Educators' Awareness of
  the Cognitive Experiences of Pupils with Dyslexia.
\newblock {\em Teach. Coll. Rec. Voice Scholarsh. Educ.}
  {\bf 2011}, {\em 113},~181--204.

\bibitem[Passig et~al.(2008)Passig, Eden, and Rosenbaum]{Passig2008}
Passig, D.; Eden, S.; Rosenbaum, V.
\newblock The impact of virtual reality on parents’ awareness of cognitive
  perceptions of a dyslectic child.
\newblock {\em Educ. Inf. Technol.} {\bf 2008}, {\em
  13},~329–344.
\newblock {\url{https://doi.org/10.1007/s10639-008-9068-6}}.

\bibitem[Wadlington et~al.(2008)Wadlington, Elliot, and Kirylo]{Wadlington2008}
Wadlington, E.; Elliot, C.; Kirylo, J.
\newblock The Dyslexia Simulation: Impact and Implications.
\newblock {\em Lit. Res. Instr.} {\bf 2008}, {\em
  47},~264–272.
\newblock {\url{https://doi.org/10.1080/19388070802300363}}.

\bibitem[Herrera et~al.(2018)Herrera, Bailenson, Weisz, Ogle, and
  Zaki]{Herrera2018}
Herrera, F.; Bailenson, J.; Weisz, E.; Ogle, E.; Zaki, J.
\newblock Building long-term empathy: A large-scale comparison of traditional
  and virtual reality perspective-taking.
\newblock {\em PLoS ONE} {\bf 2018}, {\em 13}, {0204494}.
\newblock {\url{https://doi.org/10.1371/journal.pone.0204494}}.

\bibitem[De~Luca et~al.(2023)De~Luca, Gatto, Liaci, Corchia, Chiarello,
  Faggiano, Sumerano, and De~Paolis]{de_luca_virtual_2023}
De~Luca, V.; Gatto, C.; Liaci, S.; Corchia, L.; Chiarello, S.; Faggiano, F.;
  Sumerano, G.; De~Paolis, L.T.
\newblock Virtual {Reality} and {Spatial} {Augmented} {Reality} for {Social}
  {Inclusion}: {The} {Includiamoci} {Project}.
\newblock {\em Information} {\bf 2023}, {\em 14}, {38.}
\newblock {\url{https://doi.org/10.3390/info14010038}}.

\bibitem[{Meta Platforms, Inc.}(2025)]{MetaVRForGood}
{Meta Platforms, Inc.}.
\newblock VR for Good,  2025.
\newblock {\url{https://about.meta.com/community/vr-for-good/}}.
\newblock {accessed on 12 February 2025.} 


\bibitem[Didehbani et~al.(2016)Didehbani, Allen, Kandalaft, Krawczyk, and
  Chapman]{didehbani_virtual_2016}
Didehbani, N.; Allen, T.; Kandalaft, M.; Krawczyk, D.; Chapman, S.
\newblock Virtual {Reality} {Social} {Cognition} {Training} for children with
  high functioning autism.
\newblock {\em Comput. Hum. Behav.} {\bf 2016}, {\em 62},~703--711.
\newblock {\url{https://doi.org/10.1016/j.chb.2016.04.033}}.

\bibitem[Serafin et~al.(2023)Serafin, Adjorlu, and Percy-Smith]{Serafin2023}
Serafin, S.; Adjorlu, A.; Percy-Smith, L.M.
\newblock A Review of Virtual Reality for Individuals with Hearing Impairments.
\newblock {\em Multimodal Technol. Interact.} {\bf 2023}, {\em
  7},~36.
\newblock {\url{https://doi.org/10.3390/mti7040036}}.

\bibitem[ARTE(2016)]{arteNotesOnBlindness}
ARTE.
\newblock Notes on Blindness: Into Darkness,  2016.
\newblock {Available online:} 
\url{https://www.arte.tv/digitalproductions/en/notes-on-blindness/} ({accessed on 19 February 2025).} 


\bibitem[Hoter and Nagar(2022)]{hoter_effects_2022}
Hoter, E.; Nagar, I.
\newblock The effects of a wheelchair simulation in a virtual world.
\newblock {\em Virtual Real.} {\bf 2022}, {\em 27}, {407--419.}
\newblock {\url{https://doi.org/10.1007/s10055-022-00625-7}}.

\bibitem[Rodríguez~Cano et~al.(2021)Rodríguez~Cano, Delgado~Benito,
  Casado~Muñoz, Cubo~Delgado, Ausín~Villaverde, and Santa
  Olalla~Mariscal]{rodriguez_cano_tecnologias_2021}
Rodríguez~Cano, S.; Delgado~Benito, V.; Casado~Muñoz, R.; Cubo~Delgado, E.;
  Ausín~Villaverde, V.; Santa Olalla~Mariscal, G.
\newblock Tecnologías emergentes en educación inclusiva: realidad virtual y
  realidad aumentada: {Proyecto} europeo {FORDYS}-{VAR}.
\newblock {\em Rev. Infad Psicol. Int. J. Dev. Educ. Psychol.} {\bf 2021}, {\em 2},~443--450.
\newblock {\url{https://doi.org/10.17060/ijodaep.2021.n1.v2.2093}}.

\bibitem[Pedroli et~al.(2017)Pedroli, Padula, Guala, Meardi, Riva, and
  Albani]{pedroli_psychometric_2017}
Pedroli, E.; Padula, P.; Guala, A.; Meardi, M.T.; Riva, G.; Albani, G.
\newblock A psychometric tool for a virtual reality rehabilitation approach for
  dyslexia.
\newblock {\em Comput. Math. Methods Med.} {\bf 2017},
  {\em 2017}, {7048676}.
\newblock {\url{https://doi.org/10.1155/2017/7048676}}.

\bibitem[Kaplan-Rakowski et~al.(2023)Kaplan-Rakowski, Dhimolea, and
  Khukalenko]{KaplanRakowski2023}
Kaplan-Rakowski, R.; Dhimolea, T.K.; Khukalenko, I.S.
\newblock Language teachers’ beliefs about using high-immersion virtual
  reality.
\newblock {\em Educ. Inf. Technol.} {\bf 2023}, {\em
  28},~12505--12525.
\newblock {\url{https://doi.org/10.1007/s10639-023-11686-9}}.

\bibitem[Zingoni et~al.(2023)Zingoni, Taborri, and Calabrò]{zingoni_ML_2023}
Zingoni, A.; Taborri, J.; Calabrò, G.
\newblock A machine learning-based classification model to support university
  students with dyslexia with personalized tools and strategies.
\newblock {\em Sci. Rep.} {\bf 2023}, {\emph{14}, 273.}
\newblock {\url{https://doi.org/10.21203/rs.3.rs-2783354/v1}}.

\bibitem[Alcalde-Llergo et~al.(2023)Alcalde-Llergo, Yeguas-Bolívar,
  Aparicio-Martínez, Zingoni, Taborri, and Pinzi]{AlcaldePotionsMetroxraine}
Alcalde-Llergo, J.M.; Yeguas-Bolívar, E.; Aparicio-Martínez, P.; Zingoni, A.;
  Taborri, J.; Pinzi, S.
\newblock A VR Serious Game to Increase Empathy Towards Students with
  Phonological Dyslexia.
\newblock In Proceedings of the 2023 IEEE International Conference on Metrology
  for eXtended Reality, Artificial Intelligence and Neural Engineering
  (MetroXRAINE), {Milano, Italy, 25--27 October 2023;} 
 \mbox{pp.~184--188.}
\newblock {\url{https://doi.org/10.1109/MetroXRAINE58569.2023.10405809}}.

\bibitem[Yeguas-Bolívar et~al.(2022)Yeguas-Bolívar, Alcalde-Llergo,
  Aparicio-Martínez, Taborri, Zingoni, and
  Pinzi]{yeguas-bolivar_determining_2022}
Yeguas-Bolívar, E.; Alcalde-Llergo, J.M.; Aparicio-Martínez, P.; Taborri, J.;
  Zingoni, A.; Pinzi, S.
\newblock Determining the {Difficulties} of {Students} {With} {Dyslexia} via
  {Virtual} {Reality} and {Artificial} {Intelligence}: {An} {Exploratory}
  {Analysis}.
\newblock In Proceedings of the 2022 {IEEE} {International} {Conference} on
  {Metrology} for {Extended} {Reality}, {Artificial} {Intelligence} and
  {Neural} {Engineering} ({MetroXRAINE}), {Rome, Italy, 26--28 October 2022;} pp. 585--590.
\newblock {\url{https://doi.org/10.1109/MetroXRAINE54828.2022.9967589}}.

\bibitem[Haas(2014)]{unity_2014}
Haas, J.K.
\newblock A History of the Unity Game Engine { 2014}.
\newblock Available online: \url{https://api.semanticscholar.org/CorpusID:86824974} ({accessed on 19 February 2025}).

\bibitem[Meta(2020)]{Meta}
Meta.
\newblock Meta Quest,  2020.
\newblock {Available online:} 
 \url{https://www.meta.com/en/quest/} ({accessed on 19 February 2025).}

\bibitem[Schwaber and Sutherland(2020)]{Schwaber2020}
Schwaber, K.; Sutherland, J.
\newblock {The Scrum Guide The Definitive Guide to Scrum: The Rules of the
  Game.,  2020.}
   \newblock{\url{https://scrumguides.org/scrum-guide.html} ({accessed on 19 February 2025).}}


\bibitem[Morciano et~al.(2024)Morciano, {Alcalde Llergo}, Zingoni, {Yeguas
  Bolívar}, Taborri, and Calabrò]{Morciano2024}
Morciano, G.; {Alcalde Llergo}, J.M.; Zingoni, A.; {Yeguas Bolívar}, E.;
  Taborri, J.; Calabrò, G.
\newblock Use of recommendation models to provide support to dyslexic students.
\newblock {\em Expert Syst. Appl.} {\bf 2024}, {\em 249},~123738.
\newblock {\url{https://doi.org/10.1016/j.eswa.2024.123738}}.

\bibitem[Rossi et~al.(2023)Rossi, Viola, Toni, and Cesar]{DoFRossi2023}
Rossi, S.; Viola, I.; Toni, L.; Cesar, P.
\newblock Extending 3-DoF Metrics to Model User Behaviour Similarity in 6-DoF
  Immersive Applications.
\newblock In Proceedings of the 14th ACM Multimedia Systems
  Conference, New York, NY, USA, {7--10 June 2023}; MMSys '23, pp. 39--50.
\newblock {\url{https://doi.org/10.1145/3587819.3590976}}.

\bibitem[Lukov et~al.(2015)Lukov, Friedmann, Shalev, Khentov-Kraus, Shalev,
  Lorber, and Guggenheim]{lukov_dissociations_2015}
Lukov, L.; Friedmann, N.; Shalev, L.; Khentov-Kraus, L.; Shalev, N.; Lorber,
  R.; Guggenheim, R.
\newblock Dissociations between developmental dyslexias and attention deficits.
\newblock {\em Front. Psychol.} {\bf 2015}, {\em 5}, {01501}.
\newblock {\url{https://doi.org/10.3389/fpsyg.2014.01501}}.

\bibitem[Caldani et~al.(2021)Caldani, Baghdadi, Peyre, Khoury, Delorme, and
  Bucci]{disorientationCaldani2021}
Caldani, S.; Baghdadi, M.; Peyre, H.; Khoury, E.; Delorme, R.; Bucci, M.P.
\newblock Poor visuo-spatial orientation and path memorization in children with
  dyslexia.
\newblock {\em Nord. J. Psychiatry} {\bf 2021}, {\em 76}, {195--201.}
\newblock {\url{https://doi.org/10.1080/08039488.2021.1943705}}.

\bibitem[Lie et~al.(2023)Lie, Røykenes, Sæheim, and Groven]{Lie2023}
Lie, S.S.; Røykenes, K.; Sæheim, A.; Groven, K.S.
\newblock Developing a Virtual Reality Educational Tool to Stimulate Emotions
  for Learning: Focus Group Study.
\newblock {\em Jmir Form. Res.} {\bf 2023}, {\em 7},~e41829.
\newblock {\url{https://doi.org/10.2196/41829}}.

\bibitem[{Government of Spain}(2018)]{LOPDGDD2018}
{Government of Spain}.
\newblock Organic Law on the Protection of Personal Data and Guarantee of
  Digital Rights,  2018.
\newblock {Organic Law 3/2018, of December 5.}
\newblock{\url{https://www.boe.es/eli/es/lo/2018/12/05/3/con} ({accessed on 19 February 2025).}}


\bibitem[Association(2013)]{helsinki20123}
Association, W.M.
\newblock World Medical Association Declaration of Helsinki: Ethical Principles
  for Medical Research Involving Human Subjects.
\newblock {\em JAMA} {\bf 2013}, {\em 310},~{2191--2194.} 
\newblock {\url{https://doi.org/10.1001/jama.2013.281053}}.

\bibitem[Tong et~al.(2020)Tong, Gromala, Ziabari, and Shaw]{Tong2020}
Tong, X.; Gromala, D.; Ziabari, S.P.K.; Shaw, C.D.
\newblock Designing a Virtual Reality Game for Promoting Empathy Toward
  Patients with Chronic Pain: Feasibility and Usability Study.
\newblock {\em JMIR Serious Games} {\bf 2020}, {\em 8}, {e17354.}
\newblock {\url{https://doi.org/10.2196/17354}}.

\bibitem[Bryant et~al.(2019)Bryant, Brunner, and Hemsley]{Bryant2019}
Bryant, L.; Brunner, M.; Hemsley, B.
\newblock A review of virtual reality technologies in the field of
  communication disability: implications for practice and research.
\newblock {\em Disabil. Rehabil. Assist. Technol.} {\bf 2019},
  {\em 15},~365--372.
\newblock {\url{https://doi.org/10.1080/17483107.2018.1549276}}.

\bibitem[Lin et~al.(2024)Lin, Li, Chen, and Xiong]{Lin2024}
Lin, X.; Li, R.; Chen, Z.; Xiong, J.
\newblock Design strategies for VR science and education games from an embodied
  cognition perspective: a literature-based meta-analysis.
\newblock {\em Front. Psychol.} {\bf 2024}, {\em 14}, {1292110}.
\newblock {\url{https://doi.org/10.3389/fpsyg.2023.1292110}}.

\bibitem[Jin et~al.(2023)Jin, Lee, Kim, Seo, Jung, Lim, and Lee]{Jin2023}
Jin, Y.; Lee, S.; Kim, S.; Seo, J.; Jung, K.; Lim, H.; Lee, J.
\newblock DiVRsity: Design and Development of Group Role-Play VR Platform for
  Disability Awareness Education.
\newblock In Proceedings of the 2023 ACM Designing
  Interactive Systems Conference, ACM, {Pittsburgh, PA, USA, 10--14 July 2023}; pp. 161--174.
\newblock {\url{https://doi.org/10.1145/3563657.3596047}}.

\bibitem[Rello et~al.(2016)Rello, Williams, Ali, White, and Bigham]{Rello2016}
Rello, L.; Williams, K.; Ali, A.; White, N.C.; Bigham, J.P.
\newblock Dytective: Towards Detecting Dyslexia across Languages Using an
  Online Game.
\newblock In Proceedings of the 13th International Web for
  All Conference, New York, NY, USA,  {11--13 April 2016}; W4A '16.
\newblock {\url{https://doi.org/10.1145/2899475.2899491}}.

\bibitem[Spreng* et~al.(2009)Spreng*, McKinnon*, Mar, and Levine]{Spreng2009}
Spreng, R.N.; McKinnon, M.C.; Mar, R.A.; Levine, B.
\newblock The Toronto Empathy Questionnaire: Scale Development and Initial
  Validation of a Factor-Analytic Solution to Multiple Empathy Measures.
\newblock {\em J. Personal. Assess.} {\bf 2009}, {\em
  91},~62--71.
\newblock {\url{https://doi.org/10.1080/00223890802484381}}.

\bibitem[Pérez-López et~al.(2024)Pérez-López, Navarro-Mateos, and
  Mora-Gonzalez]{PrezLpez2024}
Pérez-López, I.J.; Navarro-Mateos, C.; Mora-Gonzalez, J.
\newblock Impact of a digital serious game on emotional variables of students
  of the master’s degree in teaching.
\newblock  {\em Innov. Educ. Teach. Int.} {{\textbf{2024}}}, {\textbf{61 1}, 1--13}. 
\newblock {\url{https://doi.org/10.1080/14703297.2024.2377787}}.

\end{thebibliography}
\end{document}